\documentclass[12pt]{article}
\usepackage[utf8]{inputenc}
\usepackage[top=1in, bottom=1in, left=1in, right=1in]{geometry}
\usepackage{graphicx} 
\usepackage{subfig}
\usepackage{amsthm}
\usepackage[comma,numbers]{natbib}
\usepackage{ulem}
\usepackage{color}
\usepackage{lineno}  
\usepackage{hyperref}
\usepackage{amsmath}
\usepackage{amssymb}
\usepackage{cases}
\usepackage{bm}
\usepackage{indentfirst} 
\usepackage{mathrsfs}
\usepackage{multirow}
\usepackage{algorithm}
\usepackage{algorithmic}
\usepackage{threeparttable}
\usepackage{booktabs}
\usepackage[all]{xy}
\usepackage{hyperref}
\hypersetup{
hypertex=true,
colorlinks=true,
linkcolor=blue,
anchorcolor=blue,
citecolor=blue
}

\newcommand\keywords[1]{\textbf{Keywords}: #1}

\title{A deep learning framework for jointly solving transient Fokker-Planck equations with arbitrary parameters and initial distributions}

\author{\small{Xiaolong Wang$^{1,2,3}$, Jing Feng$^4$, Qi Liu$^5$, Chengli Tan$^{2,3}$, Yuanyuan Liu$^4$, Yong Xu$^{2,3}$\footnote{Corresponding author. E-mail addresses:  {\it hsux3@nwpu.edu.cn} (Y. Xu)}} \\
\small{$^1$ School of Mathematics and Statistics,}\\ \small{Shaanxi Normal University, Xi'an, 710119, China}\\
\small{$^2$ School of Mathematics and Statistics,}\\ \small{Northwestern Polytechnical University, Xi'an, 710129, China}\\
\small{$^3$ MOE Key Laboratory for Complexity Science in Aerospace,}\\ \small{Northwestern Polytechnical University, Xi'an, 710072, China}\\
\small{$^4$ School of Science, Xi'an University of Posts and Telecommunications,}\\ \small{Xi'an, 710121, China}\\
\small{$^5$ Department of Systems and Control Engineering, Institute of Science Tokyo,}\\ \small{Tokyo, 152-8552, Japan}
}

\date{}

\begin{document}

\maketitle
 

\begin{abstract} 
Efficiently solving the Fokker-Planck equation (FPE) is central to analyzing complex parameterized stochastic systems. However, current numerical methods lack parallel computation capabilities across varying conditions, severely limiting comprehensive parameter exploration and transient analysis. This paper introduces a deep learning-based pseudo-analytical probability solution (PAPS) that, via a single training process, simultaneously resolves transient FPE solutions for arbitrary multi-modal initial distributions, system parameters, and time points. The core idea is to unify initial, transient, and stationary distributions via Gaussian mixture distributions (GMDs) and develop a constraint-preserving autoencoder that bijectively maps constrained GMD parameters to unconstrained, low-dimensional latent representations. In this representation space, the panoramic transient dynamics across varying initial conditions and system parameters can be modeled by a single evolution network. Extensive experiments on paradigmatic systems demonstrate that the proposed PAPS maintains high accuracy while achieving inference speeds four orders of magnitude faster than GPU-accelerated Monte Carlo simulations. This efficiency leap enables previously intractable real-time parameter sweeps and systematic investigations of stochastic bifurcations. By decoupling representation learning from physics-informed transient dynamics, our work establishes a scalable paradigm for probabilistic modeling of multi-dimensional, parameterized stochastic systems.
\bigskip

\noindent\keywords{Fokker-Planck equation, transient solution, Gaussian mixture distribution, autoencoder, parallel computation.}
\end{abstract}

\section{Introduction}

As a fundamental equation in stochastic dynamics, the Fokker-Planck equation (FPE)~\cite{Risken1996FP,Frank2005Nonlinear} deterministically describes the time evolution of probability density functions (PDFs) of the system states, offering a profound link between microscopic randomness and macroscopic, predictable statistical behavior. Therefore, obtaining transient and stationary solutions of the FPE rapidly and accurately is crucial for unveiling the complex stochastic response across diverse fields, including mechanics~\cite{Xu2011Stochastic}, ecology~\cite{Wang2020Stochastic}, energy~\cite{Arenas2020Fokker}, biomedicine~\cite{Du2026High}, and neuroscience~\cite{Fagerholm2025}. However, the transient responses of nonlinear stochastic systems depend on time, as well as on multiple system parameters and the initial distributions. To comprehensively investigate the system's dynamical behaviors under varying conditions, researchers must repeatedly solve the FPE for each distinct combination of parameters and initial states. This process incurs prohibitively high computational costs, severely hindering efficient exploration of the parameter space and systematic analysis. Furthermore, analytical solutions for the FPE are confined to a few special cases and are highly nontrivial to derive~\cite{Du2026High,An2025Unique}. This makes the development of an efficient parallel numerical solver for diverse conditions central to advancing stochastic dynamical systems analysis and design.

Currently, numerical methods for solving the FPE are broadly categorized into two classes: traditional numerical methods and emerging deep learning (DL)-based approaches. Traditional methods, such as finite difference~\cite{Kumar2006Sadhana} and finite element methods~\cite{Naprstek2017Evolutionary,Georgoulis2021Hypocoercivity}, rely on spatial grid discretization and struggle to overcome the curse of dimensionality, i.e., the number of grid points increases exponentially with the system dimension. While Monte-Carlo simulations (MCS) are not limited by dimensionality, their slow convergence rate leads to poor computational efficiency. To overcome these limitations, deep learning has recently provided new avenues for obtaining FPE solutions efficiently. For instance, the DL-FP method~\cite{Xu2020Solving,Zhang2022Solving} uses mesh-free fully connected neural networks to parameterize stationary solutions, incorporating equation, initial, and boundary conditions into a single loss function. Since then, a central challenge in numerical FPE solvers is developing compact, structure preserving representations of solutions to improve the modeling of strongly nonlinear, high-dimensional systems and enhance computational tractability. Existing techniques include spectral approximation~\cite{Leonenko2009On}, polynomial chaos expansion~\cite{Zanella2020Structure}, radial basis function neural networks~\cite{Wang2025novel}, Gaussian mixture distributions (GMDs)~\cite{ER1998201,Tabandeh2022Numerical,Anderson2024Fisher}, tensor networks~\cite{Zhang2023Deep,Tang2024Solving,Wang2025Tensor}, normalizing flows~\cite{bekri2025FlowKac,He2025Adaptive,Xu2025Adaptive}, among others~\cite{Leonenko2015Numerical,Salleh2024Numerical,Sun2015numerical}. However, neither category has demonstrated the capability for parallel computation across multiple initial conditions and system parameters.

Compared to traditional numerical solvers, the core strength of DL lies in its ability to learn extensive mapping relationships during the training phase, which subsequently enables fast, parallel predictions for different tasks during inference. DL thus holds the promise of constructing a pseudo-analytical probability solution (PAPS) for the FPE. Such a solution would function similarly to a closed-form solution, yielding the corresponding probability distribution immediately for any given set of initial conditions, control parameters, and evolution time. To realize this goal, two key challenges must be addressed: (i) designing an efficient and flexible representation for diverse PDFs, and (ii) learning the temporal evolution of these PDFs under varying parameters and initial conditions in a way that supports parallel inference.

Accordingly, an ideal representation must meet two core requirements. First, it must be both general and compact. It should unify and flexibly characterize diverse PDFs from unimodal to multimodal, from broad to localized, and across the entire evolution from initial to stationary states using a low-dimensional parameterization. In contrast, traditional grid-based or dense neural network methods~\cite{Li2023Artificial} require an excessive number of parameters even for a single solution, making them unsuitable for parallel modeling of numerous transient states due to parameter explosion. Second, the representation must intrinsically preserve the mathematical constraints of the FPE solution, such as PDF normalization and non-negativity. Explicit constraint enforcement in a parallel solver would overload the loss function with penalties, complicating optimization. A structure-preserving representation~\cite{Zanella2020Structure} inherently satisfies these constraints, thereby simplifying the optimization landscape.

The recently proposed PAPS method~\cite{Wang2025Pseudo} parameterizes PDFs using a GMD and learns a decoder that transforms system parameters to the corresponding GMD parameters. It successfully achieved parallel solving for stationary solutions of multi-parameter systems, partially reflecting the aforementioned ideas. However, extending this method to transient solving introduces additional dimensions, i.e., arbitrary initial distributions and time. This makes the solution manifold extremely high-dimensional and complex. Therefore, a crucial first step is to transform the PDFs into a low-dimensional latent space, where the evolution becomes more tractable. To further reduce the model complexity, autoencoder architectures~\cite{Lee2021Parameterized,cho2024parameterized,Farenga2025Farenga,Chen2025Modeling} employ an encoder to map the initial condition and/or system parameters into a low-dimensional representation space, model the dynamics within such space, and finally use a decoder to reconstruct the solutions back to the physical space. This encode-evolve-decode paradigm, termed latent ordinary differential equations (ODEs)~\cite{Latent2019Rubanova,Coelho2024Enhancing,Nair2025Understanding}, thus presents a promising framework for parallel solving of PDEs under varying conditions. However, for the specific problem of the FPE, we must ensure that its solution always remains a valid probability distribution, not merely an arbitrary real-valued function or vectorial state. Consequently, there is a need to develop an encoding-decoding framework specifically designed for PDFs, one that inherently possesses constraint-preserving properties.

This paper proposes a novel transient PAPS (TPAPS) framework for parallel transient FPE solving. After a single training session, the model can efficiently produce transient probability density functions for any combination of initial distributions, system parameters, and time points, covering both transient and stationary dynamics. Our framework adopts a modular design that decouples representation learning from the Fokker-Planck dynamics. This decomposition breaks the grand challenge of parallel FPE solving into two tractable sub-problems: learning a universal distribution manifold and learning pure dynamics on it. This design leads to three key innovations:

1. \textbf{Decoupling probability-space representation from physical dynamics.} We learn a universal manifold of PDFs using a GMD autoencoder with a simple convexity-preserving encoder and a constraint-embedding decoder, trained independently of the FPE dynamics. Once trained, this autoencoder can represent PDFs arising from any parameters or initial conditions without retraining. This decoupling is the first of its kind for transient FPE solving.

2. \textbf{Single-network latent evolution for arbitrary time horizons with hard-encoded initial condition.} The initial condition is hard-encoded via an affine construction, eliminating its loss term. Latent dynamics are learned from the FPE residual without pre-computed trajectory or distribution data. A recursive time-leaping scheme applies the same network repeatedly over short intervals, enabling arbitrarily long horizons. Thus, a single network efficiently captures both short- and long-term evolution, with efficient training and generalization to new parameters and initial distributions.

3. \textbf{Parallel inference across parameters, initial distributions, and time points.} After a single training session, the cost of evaluating the transient PDF for any combination of system parameters, initial distribution, and time point is only one forward pass per leap step. Crucially, this cost does not scale with the number of parameter values or initial distributions, a fundamental advantage over all existing methods, e.g., MCS, PINN, and latent ODE, for which the cost grows with each new condition. This enables large-scale parameter sweeps, multi-initial-condition comparisons, and real-time bifurcation analysis that were previously infeasible.

This work constitutes a systematic extension of the PAPS framework~\cite{Wang2025Pseudo}, equipping it with full transient evolution modeling. The remainder of this paper is organized as follows. Section 2 defines the problem. Section 3 details the proposed TPAPS framework. Section 4 presents experimental results. Section 5 discusses efficiency and limitations. Section 6 concludes the paper.

\section{Problem definition}
\label{sec:2}
We consider a general $D_\text{STA}$-dimensional ($D_\text{STA}$-D) stochastic system described by the following Ito-type stochastic differential equations (SDEs)
\begin{equation}\label{eq:sde}
\dot{\mathbf{x}}(t)=\mathbf{A}(\mathbf{x},t;\mathbf{\Theta}) + \mathbf{B}(\mathbf{x},t;\mathbf{\Theta})\mathbf{\Xi}(t),
\end{equation}
where the state $\mathbf{x}(0)$ is drawn from any initial distribution $p_\text{I}(\mathbf{x})$ in a predefined initial distribution set (IDS) $\mathcal{K}$. This set comprises a series of interesting distributions used to initialize the system, thereby capturing a wide range of transient stochastic dynamical behaviors. The definition of this set in terms of GMD parameterization will be postponed until Sec.~\ref{sec:gmd}. In Eq.~(\ref{eq:sde}), $\mathbf{x}(t)=[x_{1}(t),\cdots,x_{D_\text{STA}}(t)]^\top\in\mathbb{R}^{D_\text{STA}}$ is the state vector at time $t$. The drift vector $\mathbf{A}(\mathbf{x},t;\mathbf{\Theta})=[a_{1}(\mathbf{x},t;\mathbf{\Theta}),\cdots,a_{D_\text{STA}}(\mathbf{x},t;\mathbf{\Theta})]^\top$ and the diffusion term $\mathbf{B}(\mathbf{x},t;\mathbf{\Theta})=[b_{ij}(\mathbf{x},t;\mathbf{\Theta})]_{i,j=1}^{D_\text{STA}, M}$ are controlled by $D_{\text{PAR}}$ parameters $\mathbf{\Theta}=[\theta_1,\cdots,\theta_{D_\text{PAR}}]^\top$. $\mathbf{\Xi}(t)=[\xi_{1}(t),\cdots,\xi_M(t)]^\top$ is a vector of $M$ mutually independent standard white Gaussian noises (SWGNs) with the autocorrelation functions $\mathbb{E}(\xi_{i}(t)\xi_{i}(s))=\delta(t-s), i=1,\cdots,M$, where $\mathbb{E}(\cdot)$ denotes the expectation operator.

As the stochastic system in Eq.~(\ref{eq:sde}) is a Markov process, its transient distribution $p(\mathbf{x},t;\mathbf{\Theta},p_\text{I})$ at time $t$ corresponds to the transient solution of the FPE~\cite{Risken1996FP}
\begin{equation}\label{eq:fpe}
\left\{\begin{array}{l}
\frac{\partial }{\partial t}p(\mathbf{x}, t;\bm{\Theta}, p_\text{I})=\mathcal{L}_{\text{FP}} p(\mathbf{x}, t;\bm{\Theta}, p_\text{I}),\\
p(\mathbf{x}, 0;\bm{\Theta}, p_\text{I}) = p_\text{I}(\mathbf{x}),
\end{array}\right.
\end{equation}
for any state $\mathbf{x}\in\mathbb{R}^{D_\text{STA}}$, time $t\geq 0$, suitable system parameters $\bm{\Theta}$ and initial distribution $p_\text{I}\in\mathcal{K}$, where the FP operator is defined by
\begin{equation}
\mathcal{L}_{\text{FP}}p(\mathbf{x},t;\mathbf{\Theta},p_\text{I}) = -\sum_{i=1}^{D_{\text{STA}}}\frac{\partial }{\partial x_{i}}\left[a_i(\mathbf{x};\bm{\Theta})\cdot p(\mathbf{x},t;\mathbf{\Theta},p_\text{I})\right]+\frac{1}{2}\sum_{i=1}^{D_{\text{STA}}}\sum_{j=1}^{D_{\text{STA}}}\frac{\partial^2}{\partial x_i\partial x_j}\left[d_{ij}\cdot p(\mathbf{x},t;\bm{\Theta},p_\text{I})\right],\nonumber
\end{equation}
where $d_{ij}$ is the $ij$-entry of the diffusion matrix $\mathbf{D}=\mathbf{B}(\mathbf{x},t;\mathbf{\Theta})\mathbf{B}(\mathbf{x},t;\mathbf{\Theta})^\top$. We assume that the FPE satisfies the natural boundary condition, such that
\begin{equation}\label{eq:cond_boundary}
\lim_{\|\mathbf{x}\|\rightarrow+\infty}p(\mathbf{x},t;\bm{\Theta}, p_\text{I})=0,
\end{equation}
always holds.

To facilitate the numerical study, the time domain of Eqs.~(\ref{eq:sde}) and (\ref{eq:fpe}) is confined in the interval
\begin{equation}\label{eq:time_range}
\mathcal{T}=[0,T_\text{max}],
\end{equation}
where $T_\text{max}$ is the maximal time horizon. The parameter domain $\mathcal{P}$ is a $D_\text{PAR}$-D box
\begin{equation}\label{eq:param_domain}
\mathcal{P}=\{\mathbf{\Theta}|\theta_i\in[\theta_i^\text{min},\theta_i^\text{max}],i=1,\cdots,D_\text{PAR}\}.
\end{equation}
A large enough $D_\text{STA}$-D box is further selected as the compact state domain $\mathcal{S}$
\begin{equation}\label{eq:state_domain}
\mathcal{S}=\{\mathbf{x}|x_i\in[x_i^\text{min}, x_i^\text{max}], i=1,\cdots,D_\text{STA}\},
\end{equation}
such that the probabilities of transient PDFs are always negligible outside $\mathcal{S}$ due to the natural boundary condition in Eq.~(\ref{eq:cond_boundary}). 

The goal of this paper is twofold. (1) We seek a numerical construction
\begin{equation}\label{eq:def_TPAPS}
q(\mathbf{x},t;\mathbf{\Theta},p_\text{I}):\mathcal{S} \times \mathcal{T} \times \mathcal{P} \times \mathcal{K} \rightarrow \mathbb{R},
\end{equation}
named TPAPS, to jointly model the unavailable analytical transient solutions $p(\mathbf{x},t;\mathbf{\Theta},p_\text{I})$ with all states $\mathbf{x}$, times $t$, control parameters $\mathbf{\Theta}$, and initial distributions $p_\text{I}$ in the quaternary Cartesian product $\mathcal{S} \times \mathcal{T} \times \mathcal{P} \times \mathcal{K}$. (2) We develop a learning process such that during one training session, the TPAPS $q$ can jointly approximate $p$ in $\mathcal{S} \times \mathcal{T} \times \mathcal{P} \times \mathcal{K}$. Therefore, it is unnecessary to solve the systems with different parameters or initial distributions one by one. Once the learning is completed, the TPAPS can generate transient solutions instantaneously in $\mathcal{S} \times \mathcal{T} \times \mathcal{P} \times \mathcal{K}$, significantly accelerating the exploration of transient stochastic dynamics with variable parameters and initial distributions.

As an approximation of the transient solution $p$, an effective TPAPS $q$ should satisfy several constraints, including the non-negativity condition of the PDF
\begin{equation}\label{eq:condition_nonega}
q(\mathbf{x},t;\bm{\Theta},p_\text{I})\geq 0, \quad \forall (\mathbf{x}, t, \bm{\Theta}, p_\text{I}) \in \mathcal{S} \times \mathcal{T} \times \mathcal{P} \times \mathcal{K},
\end{equation}
the normalization condition of the PDF restricted to the state domain $\mathcal{S}$
\begin{equation}\label{eq:condition_norm}
\int_{\mathcal{S}} q(\mathbf{x},t;\bm{\Theta},p_\text{I})\mathrm{d}\mathbf{x}\approx 1, \quad \text{$\forall (t, \bm{\Theta}, p_\text{I}) \in \mathcal{T} \times \mathcal{P} \times \mathcal{K}$},
\end{equation}
and the initial condition
\begin{equation}\label{eq:condition_init}
q(\mathbf{x},0;\bm{\Theta},p_\text{I}) = p_\text{I}(\mathbf{x}), \quad \forall (\mathbf{x}, \bm{\Theta}, p_\text{I}) \in \mathcal{S} \times \mathcal{P} \times \mathcal{K}.
\end{equation}
Above all, the TPAPS satisfies the FPE in Eq.~(\ref{eq:fpe})
\begin{equation}\label{eq:condition_fpe}
\frac{\partial }{\partial t}q(\mathbf{x}, t;\bm{\Theta}, p_\text{I})=\mathcal{L}_{\text{FP}} q(\mathbf{x}, t;\bm{\Theta}, p_\text{I}), \quad \forall (\mathbf{x}, t, \bm{\Theta}, p_\text{I}) \in \mathcal{S} \times \mathcal{T} \times \mathcal{P} \times \mathcal{K}.
\end{equation}

Moreover, if the system~(\ref{eq:fpe}) is ergodic, from any initial distributions $p_\text{I}$, the transient solution $p(\mathbf{x},t;\bm{\Theta},p_\text{I})$ always converges to a unique stationary solution $\lim_{t\rightarrow+\infty} p(\mathbf{x},t;\bm{\Theta},p_\text{I})=p_\text{S}(\mathbf{x};\bm{\Theta})$, which solves the stationary FPE $\mathcal{L}_{\text{FP}} p_\text{S}(\mathbf{x};\bm{\Theta})=0$.
The stationary solution $p_\text{S}$ is vital for studying the stable behaviors of stochastic systems. Therefore, for ergodic systems, we expect the asymptotic behaviors of the TPAPS align with the true stationary solutions. When $t$ is large, the TPAPS $q(\mathbf{x},t;\bm{\Theta},p_\text{I})$ initialized at any $p_\text{I}\in\mathcal{K}$ should converge to a unique PDF $\lim_{t\rightarrow+\infty}q(\mathbf{x},t;\bm{\Theta},p_\text{I})=q_\text{S}(\mathbf{x};\bm{\Theta})$, which satisfies the stationary FPE
\begin{equation}\label{eq:sfpe}
\mathcal{L}_{\text{FP}} q_\text{S}(\mathbf{x};\bm{\Theta})=0,\quad \forall (\mathbf{x}, \bm{\Theta}) \in \mathcal{S} \times \mathcal{P}.
\end{equation}

\section{The TPAPS}
\label{sec:3}
\begin{figure}[!htb]
\center{\includegraphics[width=1\textwidth]
{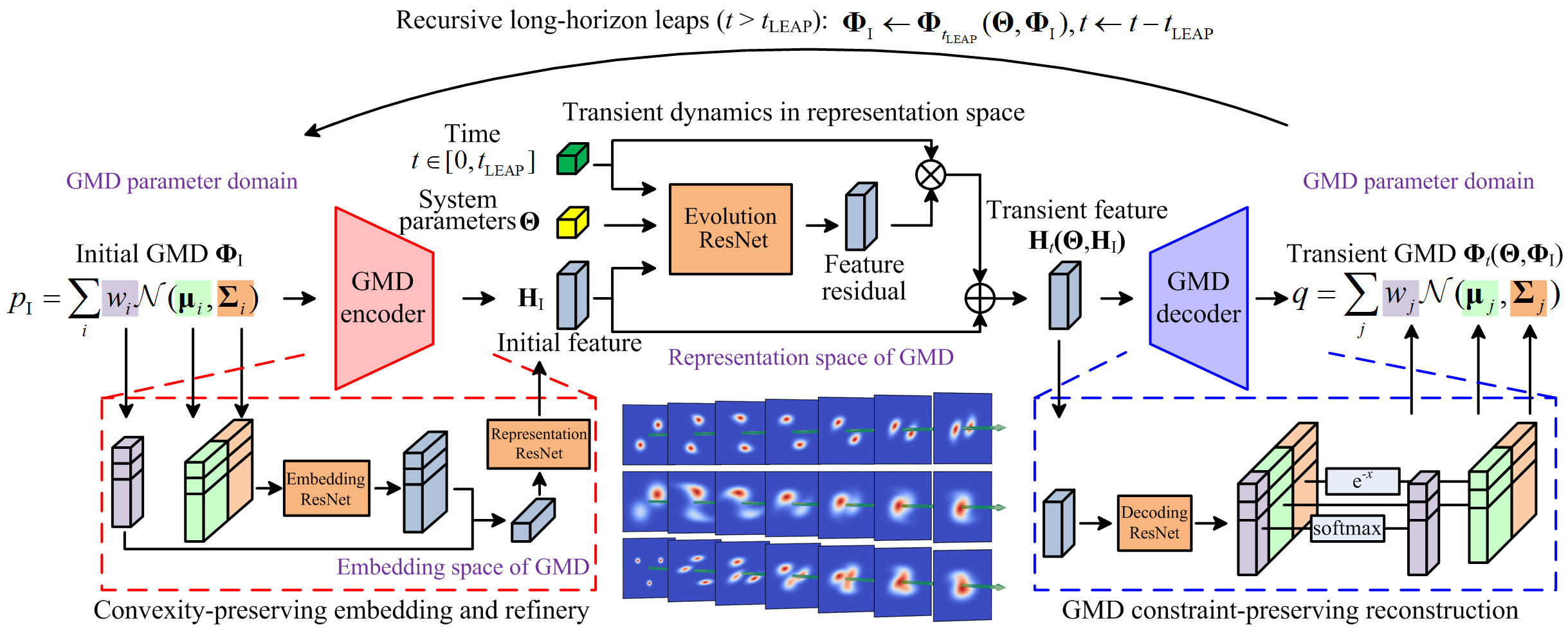}}
\caption{\label{fig:TPAPS} The architecture of the TPAPS $q(\mathbf{x},t;\bm{\Theta},p_\text{I})$ for jointly solving parameterized FPEs with diverse initial distributions and control parameters.}
\end{figure}

The transient solution $p$ on $\mathcal{S} \times \mathcal{T} \times \mathcal{P} \times \mathcal{K}$ is very intricate, not only because it has several constraints, but also because it is highly nonlinear on the system parameters and the time. To construct a capable model of the TPAPS, we perform a series of simplifications and modularizations. In Sec.~\ref{sec:gmd}, the initial, transient, and stationary distributions, as multi-dimensional multi-modal distributions, are unified and parameterized by the GMD. As the parameter space of the GMD is still high-dimensional and has several new constraints, modeling the TPAPS by tracking the evolution of the GMD parameters is computationally impractical. To this end, in Sec.~\ref{sec:gmd_ae}, a GMD autoencoder is further developed, which enables transforming GMDs into a structure-preserving low-dimensional embedding space and an unconstrained representation space. These key constructions enable us to model transient dynamics in the representation space using a single deep neural network in Sec.~\ref{sec:transient_dynamics}, which structurally satisfies the initial condition. Furthermore, the long-term transient dynamics is split into recursive short-term forecasts in Sec.~\ref{sec:recursive_calculation} to reduce the nonlinearity and alleviate the computational burden of neural networks. Finally, the network architecture is detailed in Sec.~\ref{sec:architecture} and the loss function is introduced in Sec.~\ref{sec:training}. The full architecture of the TPAPS is summarized in Fig.~\ref{fig:TPAPS}.

\subsection{The GMD}
\label{sec:gmd}

Directly modeling the TPAPS in Eq.~(\ref{eq:def_TPAPS}) with a standard neural network is inefficient. This approach requires learning a complex mapping, $q(\cdot, t;\bm{\Theta}, p_\text{I}): \mathcal{S}\rightarrow \mathbb{R}$, which must satisfy multiple constraints (Eqs.~(\ref{eq:condition_nonega})-(\ref{eq:condition_fpe})) for all times $t\in\mathcal{T}$, control parameters $\bm{\Theta}\in\mathcal{P}$, and initial distributions $p_\text{I}\in\mathcal{K}$. Enforcing all these constraints within the optimization process would lead to an intractable learning problem. Instead, we propose to model the transient distributions using the GMD. This parameterized family allows the constraints to be inherently encoded into the model structure, thereby enabling efficient and stable training.

A GMD $f_\text{GMD}(\mathbf{x};\mathbf{\Phi})$ with $n$ components is a convex combination of $n$ $D_\text{STA}$-D Gaussian distributions
\begin{equation}\label{eq:gaussian_mixture}
f_\text{GMD}(\mathbf{x};\mathbf{\Phi})=\sum_{j=1}^n w_j\cdot\mathcal{N}(\mathbf{x};\bm{\mu}_j,\mathbf{\Sigma}_j),
\end{equation}
where the weights are in the convex set
\begin{equation}\label{eq:convex_set}
\mathcal{C}(n)=\{[w_1,\cdots,w_n]^\top|\text{ All } w_j > 0 \text{ and } \sum_{j=1}^n w_j=1\}.
\end{equation}
$\mathcal{N}(\mathbf{x};\bm{\mu}_j,\mathbf{\Sigma}_j)$ is the PDF of the $j$-th Gaussian component, defined by
\begin{equation}\label{eq:gauss_distri}
\mathcal{N}(\mathbf{x};\bm{\mu}_j,\mathbf{\Sigma}_j)=\frac{1}{\sqrt{(2\pi)^{D_{\text{STA}}}|\mathbf{\Sigma}_j|}}\mathrm{e}^{-\frac{1}{2}(\mathbf{x}-\bm{\mu}_j)^\top\mathbf{\Sigma}_j^{-1}(\mathbf{x}-\bm{\mu}_j)}=
\prod_{k=1}^{D_\text{STA}}\frac{1}{\sqrt{2\pi}\sigma_{jk}}\mathrm{e}^{-\frac{(x_k-\mu_{jk})^2}{2\sigma_{jk}^2}},
\end{equation}
where $\bm{\mu}_j=[\mu_{j1},\cdots,\mu_{jD_\text{STA}}]^\top$ is the mean vector and $\mathbf{\Sigma}_j=\text{diag}\{\sigma_{j1}^2,\cdots,\sigma_{jD_{\text{STA}}}^2\}$ is the diagonal covariance matrix. 

The GMD has $n(1+2D_\text{STA})$ parameters, denoted by
\begin{equation}\label{eq:gmd_param}
\begin{split}
\mathbf{\Phi}=[\phi_1,\cdots,\phi_{n(1+2D_\text{STA})}]^\top=[w_1,\cdots,w_n,\mu_{11},\cdots,\mu_{n D_\text{STA}},\sigma_{11},\cdots,\sigma_{n D_\text{STA}}]\in\mathcal{Q}(n),
\end{split}
\end{equation}
including $n$ weights $\{w_j\}_{j=1}^n$, $n$ $D_\text{STA}$-D means $\{\bm{\mu}_j\}_{j=1}^n$, and $n D_\text{STA}$ positive standard deviations (SDs) $\{\sigma_{jk}\}_{j,k=1}^{n, D_\text{STA}}$. As a result, the feasible GMD parameter domain with $n$ components is
\begin{equation}\label{eq:gmm_param_space}
\mathcal{Q}(n)=\mathcal{C}(n)\times \mathbb{R}^{nD_\text{STA}}\times\mathbb{R}_+^{nD_\text{STA}}\subset \mathbb{R}^{n(1+2D_\text{STA})},
\end{equation}
where $\mathbb{R}_+$ is the set of positive real numbers. If $n=1$, we denote that the GMD parameters $\bm{\Phi}=(\bm{\mu}_1, \bm{\Sigma}_1)$ include the mean and the diagonal covariance matrix of the unique Gaussian distribution, whereas the single weight $w_1=1$ is omitted from $\bm{\Phi}$.

We adopt the GMD with $N_\text{GAU}$ components to represent transient distributions. The TPAPS in Eq.~(\ref{eq:def_TPAPS}) and the initial condition in Eq.~(\ref{eq:condition_init}) are expressed respectively by
\begin{equation}\label{eq:TPAPS_gmd}
\begin{split}
q(\mathbf{x}, t;\bm{\Theta}, p_\text{I})&=f_{\text{GMD}}(\mathbf{x};\bm{\Phi}_t(\bm{\Theta},\bm{\Phi}_\text{I})), \quad \forall (\mathbf{x}, t, \bm{\Theta}, \bm{\Phi}_\text{I}) \in \mathcal{S}\times\mathcal{T}\times\mathcal{P}\times \widehat{\mathcal{K}},\\
f_{\text{GMD}}(\mathbf{x};\bm{\Phi}_0(\bm{\Theta},\bm{\Phi}_\text{I}))&=f_{\text{GMD}}(\mathbf{x};\bm{\Phi}_\text{I}), \quad \forall (\mathbf{x},\bm{\Theta}, \bm{\Phi}_\text{I}) \in \mathcal{S}\times \mathcal{P}\times \widehat{\mathcal{K}}, 
\end{split}
\end{equation}
where $\bm{\Phi}_t(\bm{\Theta},\bm{\Phi}_\text{I})$ are the GMD parameters of the transient distribution at time $t$ and the initial distribution $p_\text{I}(\mathbf{x})=f_{\text{GMD}}(\mathbf{x};\bm{\Phi}_\text{I})$ is formulated by the initial GMD parameterized by $\bm{\Phi}_\text{I}\in \widehat{\mathcal{K}}$. The initial Gaussian mixture set (IGMS) $\widehat{\mathcal{K}}\subset\mathcal{Q}(N_\text{INIT})$ is a predefined set of GMD parameters with $N_\text{INIT}$ components, corresponding to the IDS $\mathcal{K}$, i.e.,
\begin{equation}
\mathcal{K}=\{f_\text{GMD}(\cdot;\bm{\Phi}_\text{I})|\bm{\Phi}_\text{I}\in\widehat{\mathcal{K}}\}.\nonumber
\end{equation}

In this work, we set $N_\text{INIT}=5$. The IGMS is defined by 
\begin{equation}\label{eq:IGMS}
\begin{split}
\widehat{\mathcal{K}}&=\{\mathbf{\Phi}_\text{I}|[w_1,\cdots,w_{N_\text{INIT}}]^\top\in \mathcal{C}(N_\text{INIT}), \mu_{jk}\in [\mu_{jk}^{\text{min}}, \mu_{jk}^{\text{max}}], \sigma_{jk}\in[\sigma_{jk}^{\text{min}}, \sigma_{jk}^{\text{max}}],\\
&j=1,\cdots,N_\text{INIT},k=1,\cdots,D_\text{STA}\},
\end{split}
\end{equation}
consisting of 5-component GMDs with restricted means and SDs. This configuration allows the IGMS to encompass test distributions that effectively have only one or two dominant Gaussian components, since some sampled weights may be close to zero. Although the corresponding IDS is limited to GMDs with at most 5 components, the convexity-preserving embedding in Sec.~\ref{sec:gmd_ae} and the recursive prediction in Sec.~\ref{sec:recursive_calculation} ensure that the resulting TPAPS remains applicable to more general cases.

By the GMD unification of initial and transient distributions, learning the TPAPS in Eq.~(\ref{eq:def_TPAPS}) is then reduced to find a transform $\bm{\Phi}_t(\bm{\Theta},\bm{\Phi}_\text{I})$ from the time $t$, the system parameters $\bm{\Theta}$ and the initial GMD parameters $\bm{\Phi}_\text{I}$ to the GMD parameters of the transient distribution
\begin{equation}\label{eq:TPAPS_gmd_transf}
\bm{\Phi}_t(\bm{\Theta},\bm{\Phi}_\text{I}):\mathcal{T} \times \mathcal{P} \times \widehat{\mathcal{K}} \subset \mathbb{R} \times \mathbb{R}^{D_\text{PAR}} \times \mathcal{Q}(N_\text{INIT}) \rightarrow \mathcal{Q}(N_\text{GAU}).
\end{equation}
This GMD parameterization has several merits. The number of variables in the GMD expression in Eq.~(\ref{eq:gmd_param}) can be precisely controlled. The non-negativity condition in Eq.~(\ref{eq:condition_nonega}) is eliminated in the learning process as the GMD automatically satisfies it. The marginal distribution of a GMD is still a GMD, which can be conveniently obtained by directly omitting the irrelevant dimensions from each component, requiring no numerical integration.

The normalization (NORM) condition for $\bm{\Phi}_t(\bm{\Theta},\bm{\Phi}_\text{I})$ in Eq.~(\ref{eq:condition_norm}), which involves a computationally expensive $D_\text{STA}$-dimensional integral, can be approximated by $D_\text{STA}$ separate normalization conditions on the 1-D marginal distributions,
\begin{equation}\label{eq:marginal_normalization}
\int_{x_k^\text{min}}^{x_k^\text{max}}\sum_{j=1}^{N_\text{GAU}}w_j\mathcal{N}(x_k;\mu_{jk},\sigma_{jk}^2)\mathrm{d}x_k\approx 1,\quad k=1,\cdots,D_\text{STA}.
\end{equation}
If we quantify the normalization conditions on a grid with $N_\text{NORM}$ points along each dimension, this reduces the number of required integral points from $O(N_\text{NORM}^{D_\text{STA}})$ to $O(N_\text{NORM} \cdot D_\text{STA})$, a crucial simplification that will be leveraged in Sec.~\ref{sec:training} to construct an efficient loss function.

\subsection{The GMD autoencoder}
\label{sec:gmd_ae}
The GMD parameter domain $\mathcal{Q}(n)$ in Eq.~(\ref{eq:gmm_param_space}) still has constraints, including the conditions that weights must form a convex combination and the SDs must be positive. These constraints complicate the modeling of the transient dynamics in Eq.~(\ref{eq:TPAPS_gmd_transf}). We now develop a GMD autoencoder to transform the constrained GMD parameters to an unconstrained feature vector and vice versa. Therefore, the transient dynamics can be easily modeled in the unconstrained representation space.

The GMD autoencoder includes an encoder and a decoder
\begin{equation}
\begin{array}{l}
\mathbf{H}=f_\text{E}(\bm{\Phi}),\\
\widehat{\bm{\Phi}}=f_\text{D}(\mathbf{H}).
\end{array}\nonumber
\end{equation}
The encoder $f_\text{E}(\cdot)$ transforms the GMD parameters $\bm{\Phi}$ with an arbitrary number of components to a feature vector $\mathbf{H}$ in the representation space, while the decoder $f_\text{D}(\cdot)$ restores the GMD parameters $\widehat{\bm{\Phi}}\in \mathcal{Q}(N_\text{GAU})$ with $N_\text{GAU}$ components from the feature $\mathbf{H}$, such that $\widehat{\bm{\Phi}}$ is functionally equivalent to $\bm{\Phi}$, i.e., $f_\text{GMD}(\mathbf{x};\widehat{\bm{\Phi}})\approx f_\text{GMD}(\mathbf{x};\bm{\Phi})$ for all $\mathbf{x}$. Note that $\widehat{\bm{\Phi}}$ and $\bm{\Phi}$ may differ significantly in the GMD parameter domain yet correspond to nearly identical distributions. For example, permuting the Gaussian components of a GMD leaves the distribution unchanged, and adding or removing components with negligible weights typically has a minor effect.

\subsubsection{The encoder}
\label{sec:gmd_ae_encoder}
Given a GMD $f_\text{GMD}(\mathbf{x};\bm{\Phi})=\sum_{i=1}^n w_{i} \mathcal{N}(\mathbf{x};\bm{\mu}_{i},\bm{\Sigma}_{i})$ with any positive integer $n$, the GMD encoder $f_\text{E}(\cdot)$ is defined by the following two-step embedding (EMB) and refined representation (REP), respectively, 
\begin{numcases}{}
\mathbf{h}=f_\text{embed}(\bm{\Phi})= \sum_{i=1}^n w_{i} \mathbf{h}_i=\sum_{i=1}^n w_{i}\cdot f_\text{EMB-RN}(\bm{\mu}_{i},\bm{\Sigma}_{i})\in\mathbb{R}^{D_\text{EMB}},\label{eq:encoder_e} \\
\mathbf{H}=f_\text{REP-RN}(\mathbf{h})\in\mathbb{R}^{D_\text{REP}}.\label{eq:encoder_r}
\end{numcases}
The embedding ResNet (EMB-RN), $f_\text{EMB-RN}$ in Eq.~(\ref{eq:encoder_e}), first transforms the $D_\text{STA}$-D mean $\bm{\mu}_i$ and the $D_\text{STA}$ SDs in the diagonal covariance matrix $\bm{\Sigma}_i$ of the $i$-th Gaussian component into a $D_\text{EMB}$-D feature vector $\mathbf{h}_i$ in the embedding space $\mathbb{R}^{D_\text{EMB}}$. These $n$ local features $\{\mathbf{h}_i\}_{i=1}^n$ are then aggregated into an integrated $D_\text{EMB}$-D feature $\mathbf{h}$ via a weighted combination using the component weights $\{w_i\}_{i=1}^n$, representing the entire GMD. Finally, $\mathbf{h}$ is refined by the representation ResNet (REP-RN), $f_\text{REP-RN}$ in Eq.~(\ref{eq:encoder_r}), to produce the final $D_\text{REP}$-D representation $\mathbf{H}$ in the refined representation space $\mathbb{R}^{D_\text{REP}}$ for downstream tasks. This second refinery stage is designed to fine-tune the embedding, ensuring the encoder's output is optimally adapted for the subsequent modeling of transient dynamics. In this work, we fix $D_\text{EMB}=50$ and $D_\text{REP}=100$. As the representation space will be used to model the transient dynamics, it is more capacious than the embedding space.

The embedding in Eq.~(\ref{eq:encoder_e}) has two promising properties to facilitate an efficient embedding of mixture distributions.

(1) Invariance of permutation symmetry of the GMD for reducing redundant encoding. We suppose that $\tau$ is an element of the permutation group $\mathcal{G}(n)$ such that $\{\tau(i)\}_{i=1}^n$ is a permutation of the indices $\{i\}_{i=1}^n$. The GMD parameters $\bm{\Phi}_\tau$ correspond to the GMD $\sum_{i=1}^nw_{\tau(i)}\mathcal{N}(\mathbf{x};\bm{\mu}_{\tau(i)},\bm{\Sigma}_{\tau(i)})$, which permutes the Gaussian components in $\bm{\Phi}$. As all the $n!$ parameter choices $\{\bm{\Phi}_\tau\}_{\tau\in \mathcal{G}(n)}$ represent the same distribution, we expect that they are encoded to a unique feature. This property is satisfied by Eq.~(\ref{eq:encoder_e}) since the summation is invariant to permutation, i.e.,
\begin{equation}
f_\text{embed}({\bm{\Phi}}_{\tau})=\sum_{i=1}^n w_{\tau(i)} f_\text{EMB-RN}(\bm{\mu}_{\tau(i)},\bm{\Sigma}_{\tau(i)}) = \sum_{i=1}^n w_i f_\text{EMB-RN}(\bm{\mu}_i,\bm{\Sigma}_i)=f_\text{embed}({\bm{\Phi}}), \quad \forall \tau\in\mathcal{G}(n).\nonumber
\end{equation}

(2) Convexity-preserving embedding for better generalization. As the GMD parameter domain in Eq.~(\ref{eq:gmm_param_space}) is still high-dimensional, the training dataset is always limited. Therefore, we expect the embedding can reflect the convex structure of the FPE, i.e., any convex combination of various transient solutions is itself a valid transient solution of the same equation, such that the autoencoder can be generalized beyond the training dataset. A natural idea is to let the embedding (EMB) operator and the convex combination (CC) of GMDs commute. To be more specific, if $\{\bm{\Phi}_i\}_{i=1}^m$ are parameters of any $m$ GMDs $\{f_\text{GMD}(\mathbf{x};\bm{\Phi}_i)\}_{i=1}^m$, the convex combination $\sum_{i=1}^m v_i f_\text{GMD}(\mathbf{x};\bm{\Phi}_i)=f_\text{GMD}(\mathbf{x};\overline{\bm{\Phi}})$ is obviously a GMD with some parameters $\overline{\bm{\Phi}}$. Then it is promising that with the same weights $\{v_i\}_{i=1}^m$, the embedding of the convex combination equals the convex combination of the embeddings.
\begin{equation}
    \xymatrix{
        \{\bm{\Phi}_i\}_{i=1}^m \ar[r]^{\text{CC}} \ar[d]^{\text{EMB}} & \overline{\mathbf{\Phi}} \ar[d]^{\text{EMB}} \\
        {\{f_\text{embed}(\bm{\Phi}_i)\}_{i=1}^m} \ar[r]^{\text{CC}} & f_\text{embed}(\overline{\bm{\Phi}}),
    }\nonumber
\end{equation}
or equivalently,
\begin{equation}\label{eq:condition_convex_1}
f_\text{embed}(\overline{\bm{\Phi}}) = \sum_{i=1}^m v_i\cdot f_\text{embed}(\bm{\Phi}_i), \quad \forall [v_1,\cdots,v_m]^\top\in\mathcal{C}(m).
\end{equation}

The property in Eq.~(\ref{eq:condition_convex_1}) not only brings computational efficiency but also grants the autoencoder strong generalization ability. Since any complex GMD can be decomposed into a combination of simpler, few-component GMDs, its embedding can be synthesized from its constituent parts. This enables our model trained on tractable, few-component GMDs to be directly applied to more challenging, multi-component ones. For instance, consider a GMD formed by mixing two simpler GMDs: $(1-\tau)f_\text{GMD}(\mathbf{x};\bm{\Phi}_1)+\tau f_\text{GMD}(\mathbf{x};\bm{\Phi}_2)$, where $\bm{\Phi}_1$ and $\bm{\Phi}_2$ have $k_1$ and $k_2$ components respectively. Although this mixture contains $k_1+k_2$ components, a structure unseen during training, its embedding is simply the same convex combination of the individual embeddings.

Essentially, Eq.~(\ref{eq:encoder_e}) and the more general Eq.~(\ref{eq:condition_convex_1}) imply each other, establishing their equivalence. If we confine that each GMD in Eq.~(\ref{eq:condition_convex_1}) consists of a single Gaussian distribution, it reduces to Eq.~(\ref{eq:encoder_e}). On the contrary, if we define $f_\text{GMD}(\mathbf{x};\bm{\Phi}_i)=\sum_{j=1}^n w_{ij} \mathcal{N}(\mathbf{x};\bm{\mu}_{ij},\bm{\Sigma}_{ij})$, and $f_\text{GMD}(\mathbf{x};\overline{\bm{\Phi}})=\sum_{i=1}^m\sum_{j=1}^n v_iw_{ij} \mathcal{N}(\mathbf{x};\bm{\mu}_{ij},\bm{\Sigma}_{ij})$, Eq.~(\ref{eq:encoder_e}) implies Eq.~(\ref{eq:condition_convex_1}) as follows
\begin{equation}
f_\text{embed}(\overline{\bm{\Phi}}) = \sum_{i=1}^m\sum_{j=1}^n v_iw_{ij} \cdot f_\text{EMB-RN}(\bm{\mu}_{ij},\bm{\Sigma}_{ij}) = \sum_{i=1}^m v_i \cdot f_\text{embed}(\bm{\Phi}_i).\nonumber
\end{equation}
Therefore, up to the specific parameterization of the ResNet, Eq.~(\ref{eq:encoder_e}) represents the unique construction that satisfies both properties.

\subsubsection{The decoder}
\label{sec:gmd_ae_decoder}
The structure of the decoder functions by mapping a vectorial representation back into the GMD parameter domain. Given a feature representation $\mathbf{H}\in\mathbb{R}^{D_\text{REP}}$, the GMD decoder $f_\text{D}$ restores the GMD parameters $\widehat{\bm{\Theta}}\in\mathcal{Q}(N_\text{GAU})$ through two successive functions
\begin{numcases}
\quad\mathbf{H}_D = f_\text{DEC-RN}(\mathbf{H})\in\mathbb{R}^{N_\text{GAU}(1+2D_\text{STA})},\\
\widehat{\bm{\Theta}} = f_\text{SUR}(\mathbf{H}_D)\in\mathcal{Q}(N_\text{GAU}).
\end{numcases}
Serving as a goal-oriented shaping network, the decoding ResNet (DEC-RN) $f_\text{DEC-RN}$ molds the feature $\mathbf{H}$ into a target-specific vector $\mathbf{H}_D=[h_1,\cdots,h_{N_\text{GAU}(1+2D_\text{STA})}]^\top$ within an unconstrained space of dimension $N_\text{GAU}(1+2D_\text{STA})$. The surjection $f_\text{SUR}$ then maps the intermediate vector $\mathbf{H}_D$ to the GMD parameters $\widehat{\mathbf{\Phi}}=[\hat{\phi}_1,\cdots,\hat{\phi}_{N_\text{GAU}(1+2D_\text{STA})}]^\top\in\mathcal{Q}(N_\text{GAU})$ with $N_\text{GAU}$ components,
\begin{equation}\label{eq:surjection}
\hat{\phi}_k=\left\{\begin{array}{ll}
\text{e}^{h_k}/(\sum_{k=1}^{N_\text{GAU}}\text{e}^{h_k}),&1\leq k\leq N_\text{GAU},\\
h_k,&N_\text{GAU}+1\leq k \leq N_\text{GAU}(1+D_\text{STA}),\\
\text{e}^{-h_k},&N_\text{GAU}(1+D_\text{STA})+1\leq k \leq N_\text{GAU}(1+2D_\text{STA}).
\end{array}\right.
\end{equation}
The first part is the softmax function that maps the first $N_\text{GAU}$ elements of $\mathbf{H}_D$ onto the positive convex set $\mathcal{C}(N_\text{GAU})$ in Eq.~(\ref{eq:convex_set}). The second part takes the intermediate $N_\text{GAU}D_\text{STA}$ real elements of $\mathbf{H}_D$ as the $N_\text{GAU}$ mean vectors. The third part transforms the last $N_\text{GAU}D_\text{STA}$ elements to the positive SDs required by the $N_\text{GAU}$ $D_\text{STA}$-D diagonal covariance matrices by the function $\text{e}^{-x}$. 

This design delivers two critical advantages. It ensures that structural GMD constraints, such as the weight normalization and positive standard deviations, are built into the decoder, obviating explicit penalty terms. Furthermore, it guarantees representational capacity. Since Eq.~(\ref{eq:surjection}) is surjective from $\mathbb{R}^{N_\text{GAU}(1+2D_\text{STA})}$ to $\mathcal{Q}(N_\text{GAU})$~\cite{Wang2025Pseudo}, training $f_\text{DEC-RN}$ appropriately enables the decoder to reconstruct any desired $N_\text{GAU}$-component GMDs, such as specialized transient and stationary distributions. The representational power can be scaled by employing a deeper DEC-RN.

\subsection{Modeling transient dynamics in representation space}
\label{sec:transient_dynamics}
The GMD autoencoder establishes a bijection between the constrained GMD parameter domain and an unconstrained $D_\text{REP}$-D representation space, enabling us to model transient stochastic dynamics directly in the representation space without worrying about the constraints of the GMD in Eq.~(\ref{eq:gmm_param_space}).

The initial GMD parameters $\bm{\Phi}_\text{I}\in \widehat{\mathcal{K}}$ in Eq.~(\ref{eq:TPAPS_gmd}) are represented by an unconstrained initial feature $\mathbf{H}_\text{I}$ by the encoder
\begin{equation}\label{eq:HI}
\mathbf{H}_\text{I}=f_\text{E}(\bm{\Phi}_\text{I})\in f_\text{E}(\widehat{\mathcal{K}})\subset\mathbb{R}^{D_\text{REP}}.
\end{equation}
Crucially, the transient distribution $q(\mathbf{x},t;\bm{\Theta},p_\text{I})$ expressed by the constrained GMD parameters $\bm{\Phi}_t(\bm{\Theta}, \bm{\Phi}_\text{I})$ in Eq.~(\ref{eq:TPAPS_gmd_transf}) can be represented by an unconstrained transient feature $\mathbf{H}_t(\bm{\Theta}, \mathbf{H}_\text{I})\in\mathbb{R}^{D_\text{REP}}$. The corresponding GMD parameters are obtained by decoding the transient feature,
\begin{equation}\label{eq:Phi_t_d}
\bm{\Phi}_t(\bm{\Theta}, \bm{\Phi}_\text{I})=f_\text{D}(\mathbf{H}_t(\bm{\Theta}, \mathbf{H}_\text{I})),\quad \forall(t, \bm{\Theta}, \mathbf{H}_\text{I}) \in \mathcal{T}\times\mathcal{P}\times f_\text{E}(\widehat{\mathcal{K}}).
\end{equation}
As a result, the construction of the TPAPS is reduced to find a suitable feature transformation
\begin{equation}
\mathbf{H}_t(\bm{\Theta}, \mathbf{H}_\text{I}): \mathcal{T}\times\mathcal{P}\times f_\text{E}(\widehat{\mathcal{K}})\subset \mathbb{R}\times\mathbb{R}^{D_\text{PAR}}\times \mathbb{R}^{D_\text{REP}}\rightarrow \mathbb{R}^{D_\text{REP}},
\end{equation}
that after decoding by Eq.~(\ref{eq:Phi_t_d}), the corresponding GMD always satisfies the FPE in Eq.~(\ref{eq:fpe}) and the initial condition in Eq.~(\ref{eq:TPAPS_gmd}).

To begin constructing the latent feature $\mathbf{H}_t(\bm{\Theta}, \mathbf{H}_\text{I})$, the initial condition in Eq.~(\ref{eq:TPAPS_gmd}) is replaced by the following initial condition in the representation space
\begin{equation}\label{eq:condition_init_embed}
\mathbf{H}_0(\bm{\Theta}, \mathbf{H}_\text{I})=\mathbf{H}_\text{I},\quad \forall (\bm{\Theta}, \mathbf{H}_\text{I}) \in \mathcal{P}\times f_\text{E}(\widehat{\mathcal{K}}).
\end{equation}
By using the definitions in Eqs. (\ref{eq:HI}) and (\ref{eq:Phi_t_d}), Eq.~(\ref{eq:condition_init_embed}) ensures that the original initial condition in Eq.~(\ref{eq:TPAPS_gmd}) is satisfied, as
\begin{equation}
\bm{\Phi}_0(\bm{\Theta}, \bm{\Phi}_\text{I})=f_\text{D}(\mathbf{H}_0(\bm{\Theta}, \mathbf{H}_\text{I}))=f_\text{D}(\mathbf{H}_\text{I})=
f_\text{D}(f_\text{E}(\bm{\Phi}_\text{I}))=\widehat{\bm{\Phi}}_\text{I},
\end{equation}
and the decoded GMD parameters $\widehat{\bm{\Phi}}_\text{I}$ function identically to $\bm{\Phi}_\text{I}$.

Naturally, the evolution of vectorial features can be captured by ODEs. Inspired by the latent ODEs~\cite{Latent2019Rubanova,iakovlev2023latent}, we assume that the transient feature $\mathbf{H}_t(\bm{\Theta}, \mathbf{H}_\text{I})$ in the representation space is governed by the ODEs
\begin{equation}
\left\{
\begin{split}
\frac{\mathrm{d}}{\mathrm{d} t}\mathbf{H}_t(\bm{\Theta}, \mathbf{H}_\text{I})&=\mathbf{V}(\mathbf{H}_t(\bm{\Theta}, \mathbf{H}_\text{I}),\bm{\Theta}, t),\\
\mathbf{H}_0(\bm{\Theta}, \mathbf{H}_\text{I})&=\mathbf{H}_\text{I},
\end{split}\right.
\end{equation}
where $\mathbf{V}$ denotes the unknown latent vector fields. 

Recent reduced-order models~\cite{Chen2025Modeling,Conti2026VENI,Lee2021Parameterized,Farenga2025Farenga} learn such vector fields from a large number of pre-computed PDE snapshots. However, generating these snapshots for multi-parameter systems is computationally prohibitive, as it requires solving the governing PDE repeatedly for each parameter instance and each time step. On the other hand, if we aim at building a snapshot-free latent dynamics model solely through PDE residual minimization, the numerical integration error of the latent ODE can become unacceptably large due to the absence of ground-truth trajectories. Instead, we propose an evolution ResNet (EVO-RN) $f_\text{EVO-RN}$ to approximate the time-averaged integral of the latent vector field over the interval $[0,t]$
\begin{equation}
\frac{1}{t}\int_0^t\mathbf{V}(\mathbf{H}_s,\bm{\Theta}, s)\mathrm{d}s=\frac{1}{t}[\mathbf{H}_t(\bm{\Theta}, \mathbf{H}_\text{I}) - \mathbf{H}_\text{I}]\triangleq f_\text{EVO-RN}(t, \bm{\Theta},\mathbf{H}_\text{I}).
\end{equation} 
This motivates an affine construction of the transient feature $\mathbf{H}_t(\bm{\Theta}, \mathbf{H}_\text{I})$ as follows
\begin{equation}\label{eq:Ht}
\mathbf{H}_t(\bm{\Theta}, \mathbf{H}_\text{I})=\mathbf{H}_\text{I} + t\cdot f_\text{EVO-RN}(t, \bm{\Theta},\mathbf{H}_\text{I}),\quad \forall(t, \bm{\Theta}, \mathbf{H}_\text{I}) \in \mathcal{T}\times\mathcal{P}\times f_\text{E}(\widehat{\mathcal{K}}).
\end{equation}
Consequently, short-term transient dynamics can be captured in a single computation, eliminating the need for numerous finite-difference iterations. Combining Eqs.~(\ref{eq:HI}), (\ref{eq:Phi_t_d}), and (\ref{eq:Ht}), the transient solution in the GMD parameter domain is
\begin{equation}\label{eq:tran_dyna}
\bm{\Phi}_t(\bm{\Theta},\bm{\Phi}_\text{I}) = f_\text{D}(\mathbf{H}_t(\bm{\Theta},f_\text{E}(\bm{\Phi}_\text{I}))) = f_\text{D}(f_\text{E}(\bm{\Phi}_\text{I}) + t\cdot f_\text{EVO-RN}(t, \bm{\Theta},f_\text{E}(\bm{\Phi}_\text{I}))),\quad (t, \bm{\Theta}, \bm{\Phi}_\text{I}) \in \mathcal{T}\times\mathcal{P}\times \widehat{\mathcal{K}}.
\end{equation}

The formulation in Eq.~(\ref{eq:Ht}) provides a unified framework for modeling transient dynamics across different time scales. From a Taylor-expansion viewpoint, the latent evolution can be written as
\begin{equation}
\mathbf{H}_t(\bm{\Theta}, \mathbf{H}_\text{I})=\mathbf{H}_\text{I} + t\cdot f_\text{EVO-RN}(0, \bm{\Theta},\mathbf{H}_\text{I})+o(t),
\end{equation}
where the network output $f_\text{EVO-RN}(0, \bm{\Theta},\mathbf{H}_\text{I})$ approximates the infinitesimal generator, ensuring an affine update that maintains temporal smoothness for small $t$. As $t$ increases, the same network adaptively learns the integrated effect of higher-order terms through its time-dependent argument, thereby capturing nonlinear long-term dependencies without explicit derivative computations. Thus, a single architecture seamlessly bridges the infinitesimal regime and finite-time nonlinear evolution.

Furthermore, the proposed EVO-RN offers three interrelated advantages for parallel solving transient FPEs. First, the affine structure in Eq.~(\ref{eq:Ht}) inherently enforces the initial condition in Eq.~(\ref{eq:condition_init_embed}), eliminating a common source of bias and computation burden for parallel learning diverse initial conditions. Second, this same structure acts as a strong architectural regularizer. It confines learning to the residual term $\mathbf{H}_t-\mathbf{H}_\text{I}=t\cdot f_\text{EVO-RN}$, which naturally yields smooth, physically plausible trajectories and avoids unphysical temporal oscillations, which is similar to ResNets stabilizing deep function approximation. Third, the resulting explicit mapping from $(t, \bm{\Theta}, \mathbf{H}_\text{I})$ to $\mathbf{H}_t$ by a single neural network enables efficient parallel computation. Predictions for diverse initial distributions, system parameters, and time points can be evaluated simultaneously in batches, bypassing the sequential time-stepping of conventional PDE solvers. Together, these features yield a framework that is physically consistent, stable to train, and scalable in practice.

\subsection{Recursive calculation}
\label{sec:recursive_calculation}
We aim for the TPAPS to compute transient solutions for any time in $\mathcal{T}=[0, T_\text{max}]$. However, when $T_\text{max}$ is large, the transient dynamics exhibit extreme complexity and strong nonlinearity across different time scales. For instance, if the initial distribution is a sharply peaked Gaussian, the transient solution near $t=0$ displays localized behavior characterized by rapid peak decay and variance growth. In contrast, at large $t$, transient solutions originating from diverse initial distributions converge toward a common global steady state. This sharp dichotomy between fast, localized short-term dynamics and slow, global long-term relaxation makes it exceedingly difficult for a single neural network to accurately capture both regimes simultaneously.

To reduce the nonlinear effects and ease the learning of neural networks, the long-term calculation of the TPAPS for $t \in \mathcal{T}=[0, T_\text{max}]$ is realized by recursive short-term calculations within $t \in \widehat{\mathcal{T}}=[0, t_\text{LEAP}]$, where $t_\text{LEAP}$ is a predefined small leap time. If $t>t_\text{LEAP}$, we use the following recursive expressions to reduce the time horizon of the transient dynamics of Eq.~(\ref{eq:tran_dyna}),
\begin{equation}\label{eq:recursive}
\bm{\Phi}_t(\bm{\Theta},\bm{\Phi}_\text{I}) \triangleq \bm{\Phi}_{t-t_\text{LEAP}}(\bm{\Theta},\bm{\Phi}_{t_\text{LEAP}}(\bm{\Theta},\bm{\Phi}_\text{I})).
\end{equation}
This implies that the calculation at time $t$ is decomposed into two sequential steps at $t_\text{LEAP}$ and $t - t_\text{LEAP}$, where the latter may itself be further decomposed into multiple sub-steps. In general, for the time $t = k \cdot t_\text{LEAP} + s$ where $s \in (0, t_\text{LEAP}]$ and $k\in\mathbb{N}$, the encode-evolve-decode pipeline is executed $k+1$ times, which is summarized in Algorithm~\ref{alg:tpaps}. Specifically, the first $k$ executions each predict the dynamics over the next $t_\text{LEAP}$ time units, and the final execution covers the remaining interval of length $s\leq t_\text{LEAP}$. This recursive setting effectively shortens the training time horizon of the TPAPS from $\mathcal{T}$ to $\widehat{\mathcal{T}}$. As a result, the computational complexity of the EVO-RN is reduced, owing to the enhanced linearity of Eq.~(\ref{eq:Ht}) for small time intervals. Combining Eqs.~(\ref{eq:TPAPS_gmd}), (\ref{eq:tran_dyna}), and (\ref{eq:recursive}), the full expression of the TPAPS is
\begin{equation}\label{eq:TPAPS_full}
\begin{split}
q(\mathbf{x}, t;\bm{\Theta}, \bm{\Phi}_\text{I}) &= f_\text{GMD}(\mathbf{x};\bm{\Phi}_t(\bm{\Theta},\bm{\Phi}_\text{I})),\quad (\mathbf{x}, t, \bm{\Theta}, \bm{\Phi}_\text{I}) \in \mathcal{S}\times\mathcal{T}\times\mathcal{P}\times \widehat{\mathcal{K}},\\
\bm{\Phi}_t(\bm{\Theta},\bm{\Phi}_\text{I}) & = \left\{\begin{array}{ll}
f_\text{D}(f_\text{E}(\bm{\Phi}_\text{I}) + t\cdot f_\text{EVO-RN}(t, \bm{\Theta},f_\text{E}(\bm{\Phi}_\text{I}))), & t \in [0, t_\text{LEAP}],\\
\bm{\Phi}_{t-t_\text{LEAP}}(\bm{\Theta},\bm{\Phi}_{t_\text{LEAP}}(\bm{\Theta},\bm{\Phi}_\text{I})), & t > t_\text{LEAP}.
\end{array}\right.
\end{split}
\end{equation}

\begin{algorithm*}
\caption{Calculating transient GMD by the TPAPS}\label{alg:tpaps}
\begin{algorithmic}[1]
\REQUIRE The TPAPS (including a GMD encoder $f_\text{E}$, decoder $f_\text{D}$, and the EVO-RN $f_\text{EVO-RN}$), system parameters $\bm{\Theta}\in\mathcal{P}$, initial GMD $\bm{\Phi}_\text{I}\in\widehat{\mathcal{K}}$, time $t$.
\STATE Set $\bm{\Phi}\leftarrow\bm{\Phi}_\text{I}$ and decompose $t=k\cdot t_\text{LEAP} + s$ for $k\in \mathbb{N}$ and $s\in (0, t_\text{LEAP}]$.
\FOR {$i=1$ to $k$}
	\STATE $\bm{\Phi} \leftarrow f_\text{D}(f_\text{E}(\bm{\Phi}) + t_\text{LEAP} \cdot f_\text{EVO-RN}(t_\text{LEAP}, \bm{\Theta},f_\text{E}(\bm{\Phi})))$.
\ENDFOR
\STATE Set $\bm{\Phi}_t(\bm{\Theta},\bm{\Phi}_\text{I}) \leftarrow f_\text{D}(f_\text{E}(\bm{\Phi}) + s \cdot f_\text{EVO-RN}(s, \bm{\Theta},f_\text{E}(\bm{\Phi})))$.
\RETURN $\bm{\Phi}_t(\bm{\Theta},\bm{\Phi}_\text{I})$
\end{algorithmic}
\end{algorithm*}

As a trade-off, recursive calculation enlarges the feasible initial GMDs, which should not only contain GMDs in the original IGMS at $t=0$, but also include the intermediate transient GMDs in the recursive process. To this end, we define the transient Gaussian mixture set (TGMS) $\widehat{\mathcal{K}}(T_\text{INIT})$ for collecting the transient GMDs with a predefined maximal time $T_\text{INIT}$, where
\begin{equation}\label{eq:tgms}
\widehat{\mathcal{K}}(t) = \{\bm{\Phi}_s(\bm{\Theta},\bm{\Phi}_\text{I})|(s, \bm{\Theta}, \bm{\Phi}_\text{I})\in (0,t]\times \mathcal{P}\times\widehat{\mathcal{K}}\}.
\end{equation}
All the possible transient distributions with the time $t\in(0, T_\text{INIT}]$, parameters in the parameter domain $\mathcal{P}$ and initial GMD in the IGMS $\widehat{\mathcal{K}}$ are included in $\widehat{\mathcal{K}}(T_\text{INIT})$ in the training stage. Note that the TGMS $\widehat{\mathcal{K}}(t)$ expands monotonically with the increase of $t$, and the excluded case $t=0$ corresponds to the IGMS.

Accordingly, the recursive calculation simplifies the training domain of the TPAPS in Eq.~(\ref{eq:TPAPS_full}) from $\mathcal{S}\times\mathcal{T}\times\mathcal{P}\times \widehat{\mathcal{K}}$ to $\mathcal{S}\times\widehat{\mathcal{T}}\times\mathcal{P}\times [\widehat{\mathcal{K}}\cup\widehat{\mathcal{K}}(T_\text{INIT})]$, where the set of initial GMDs consists of both the original IGMS and the TGMS. As a result, the expected time horizon of the TPAPS for accurate prediction is $T_\text{max}=T_\text{INIT}+t_\text{LEAP}$.  Nevertheless, for ergodic systems, a moderate $T_\text{INIT}$ in the training process may still achieve long-term calculations since the transient distributions would quickly converge to the stationary distributions.

A proper training of the GMD autoencoder requires sampling the TGMS in Eq.~(\ref{eq:tgms}). However, a key difficulty stems from a circular dependency: sampling requires computing $\bm{\Phi}_t(\bm{\Theta},\bm{\Phi}_\text{I})$ via the TPAPS in Eq.~(\ref{eq:TPAPS_full}), yet the TPAPS itself is the learning objective and thus unavailable a priori.

We resolve this circularity by a collaborative bootstrapping training strategy. Throughout training, we continually use the current, evolving TPAPS to generate candidate GMDs for the TGMS, treating them as if they were true transient distributions. Initially, this yields a pool of approximate distributions. As training progresses, the GMD autoencoder improves, first accurately representing and restoring distributions with $t$ close to zero, since these are well approximated by the initial GMDs in the IGMS that the autoencoder can already handle reliably. This initial accuracy then propagates. The improved TPAPS can model distributions over slightly longer time horizons, which in turn provides higher-quality samples for the TGMS at larger $t$. Iteratively, the model's effective time horizon expands, and the GMD encoder gradually learns to represent a broad spectrum of transient distributions arising from diverse initial conditions and system parameters. Consequently, the TGMS is progressively populated with increasingly accurate transient distributions, ultimately enabling the TPAPS to learn across the full target temporal domain.

\subsection{Network architecture}
\label{sec:architecture}
The TPAPS consists of four ResNets, including the EMB-RN, REP-RN, and DEC-RN in the GMD autoencoder, as well as the EVO-RN that models transient dynamics. Each ResNet~\cite{He2015DeepRL,Wang2024Deep,Wang2025Pseudo} consists of a cascade of residual blocks (RBs). Every residual block is defined by
\begin{equation}
\mathbf{y}=u(l_3(u(l_2(u(l_1(\mathbf{x})))))) + r(\mathbf{x}),
\end{equation}
which contains three linear layers (LL) $\{l_i\}_{i=1}^3$ of the sizes $(D_\text{In}, D_\text{Mid})$, $(D_\text{Mid}, D_\text{Mid})$, and $(D_\text{Mid}, D_\text{Out})$, each followed by a Tanh activation $u(x)=(\text{e}^x - \text{e}^{-x})/(\text{e}^x + \text{e}^{-x})$. An $(m, n)$-linear layer is an affine function
\begin{equation}
l(\mathbf{x})=\mathbf{A}\mathbf{x}+\mathbf{b},
\end{equation}
which takes an $m$-D vector as input and outputs an $n$-D vector, where $\mathbf{A}$ is an $n\times m$ matrix and $\mathbf{b}$ is an $n$-D bias vector. The output of the RB is formed by a vector sum of the block's input and the three-layer transformed signal. If the input $\mathbf{x}$ and the output $\mathbf{y}$ have the same dimension, the sum is taken directly, i.e., $r(\mathbf{x})=\mathbf{x}$. Otherwise, $r(\mathbf{x})$ represents an extra linear layer applying to $\mathbf{x}$ to align dimensions before summation.

The sizes of the four ResNets are detailed in Tab.~\ref{tab:architecture}. For simplicity, we use identical neural network architectures in all numerical experiments except for necessary adaptation, such as the system dimension $D_\text{STA}$ and the number of control parameters $D_\text{PAR}$. The GMD includes $N_\text{GAU}=100$ adaptive Gaussian components. The dimension of the embedding space is $D_\text{EMB}=50$ and the dimension of the GMD representation space is $D_\text{REP}=100$. The EMB-RN $f_\text{EMB-RN}$ includes 3 RBs with 50 neurons in each linear layer. The REP-RN $f_\text{REP-RN}$ includes 3 RBs with 100 neurons in each linear layer. For modeling the complex transient dynamics and decoding tasks, the EVO-RN $f_\text{EVO-RN}$ and DEC-RN $f_\text{DEC-RN}$ include 6 RBs with 100 neurons in each linear layer. The DEC-RN has a final linear layer with input dimension 100 and output dimension $N_\text{GAU}(1+2D_\text{STA})$. It transforms the transient feature to a $N_\text{GAU}(1+2D_\text{STA})$-D vector, such that the surjection in Eq.~(\ref{eq:surjection}) can finally recover the GMD parameters.

\begin{table}[!htb]
\scriptsize 
\centering
\caption{The architectures of the ResNets in the TPAPS}\label{tab:architecture}
\begin{tabular}{ccccc}
\toprule
ResNet ID & EMB-RN & REP-RN & EVO-RN & DEC-RN\\
\midrule
Input & $\{(\mu_{ij},\sigma_{ij})\}_{i=1}^{D_\text{STA}}$ & Embedding $\mathbf{h}$ & $(t,\bm{\Theta}, \mathbf{H}_\text{I})$ & Transient $\mathbf{H}_t(\bm{\Theta}, \mathbf{H}_\text{I})$\\
$D_\text{In}$/$D_\text{Mid}$/$D_\text{Out}$ of RB\#1 & $2D_\text{STA}$/50/50 & $D_\text{EMB}$/100/100 & $1+D_\text{PAR}+D_\text{REP}$/100/100 & $D_\text{REP}$/100/100\\
$D_\text{In}$/$D_\text{Mid}$/$D_\text{Out}$ of RB\#2 & 50/50/50 & 100/100/100 & 100/100/100 & 100/100/100\\
$D_\text{In}$/$D_\text{Mid}$/$D_\text{Out}$ of RB\#3 & 50/50/$D_\text{EMB}$ & 100/100/$D_\text{REP}$ & 100/100/100 & 100/100/100\\
$D_\text{In}$/$D_\text{Mid}$/$D_\text{Out}$ of RB\#4 & & & 100/100/100 & 100/100/100\\
$D_\text{In}$/$D_\text{Mid}$/$D_\text{Out}$ of RB\#5 & & & 100/100/100 & 100/100/100\\
$D_\text{In}$/$D_\text{Mid}$/$D_\text{Out}$ of RB\#6 & & & 100/100/$D_\text{REP}$ & 100/100/100\\
In/out DIM of final LL &  &  &  & 100/$N_\text{GAU}(1+2D_\text{STA})$ \\
Output & Embedding $\mathbf{h}_j$ & Initial $\mathbf{H}_\text{I}$ & $[\mathbf{H}_t(\bm{\Theta}, \mathbf{H}_\text{I}) - \mathbf{H}_\text{I}]/t$ & Decoded feature $\mathbf{H}_\text{D}$\\
\bottomrule
\end{tabular}
\end{table}

\subsection{Network training}
\label{sec:training}
The learning of the TPAPS includes two interdependent goals. (1) By combining the EMB-RN, REP-RN, and DEC-RN, the GMD autoencoder should restore all possible initial GMDs in the IGMS $\widehat{\mathcal{K}}$ and intermediate GMDs in the TGMS $\widehat{\mathcal{K}}(T_\text{INIT})$, leading to the autoencoder loss $L_\text{AE}$. (2) In the representation space, the EVO-RN should fit short-term transient dynamics in $\widehat{\mathcal{T}}\times\mathcal{P}\times [\widehat{\mathcal{K}}\cup\widehat{\mathcal{K}}(T_\text{INIT})]$ at all states in $\mathcal{S}$, such that the TPAPS recursively solves all the FPEs in $\mathcal{S}\times\mathcal{T}\times\mathcal{P}\times \widehat{\mathcal{K}}$, leading to the FPE loss $L_\text{FP}$. 

\subsubsection{Sampling strategy}
A sufficient training of the TPAPS requires sampling the parameterized high-dimensional set $\mathcal{S}\times\widehat{\mathcal{T}}\times\mathcal{P}\times [\widehat{\mathcal{K}}\cup\widehat{\mathcal{K}}(T_\text{INIT})]$. Nevertheless, this process is straightforward, because the loss function is designed to exclude the complex initial condition and GMD constraints, as these are inherently encoded within the network architecture itself. Moreover, the sets of states, times, system parameters, and initial GMDs are mutually decoupled. Consequently, their Cartesian product can be randomly sampled by performing independent draws from each set.

We define the operator $\mathbb{U}(\mathcal{A})$ to be a single uniform random draw from the parameterized set $\mathcal{A}$. Furthermore, $\mathbb{U}(\mathcal{A}, n)$ denotes $n$ independent and uniform draws from $\mathcal{A}$. This notation naturally extends to Cartesian products. The Cartesian pairing operator $\overline{\parallel}$ is defined by
\begin{equation}
\mathbb{U}(\mathcal{A} \times \mathcal{B}, n)=\mathbb{U}(\mathcal{A}, n) \overline{\parallel} \mathbb{U}(\mathcal{B}, n),\nonumber
\end{equation}
yielding $n$ pairs $\{(a_i, b_i)\}_{i=1}^n$, where each $a_i$ and $b_i$ are drawn independently and uniformly from $\mathcal{A}$ and $\mathcal{B}$, respectively.

An initial GMD sample $\mathbb{U}(\widehat{\mathcal{K}})$, drawn from the IGMS $\widehat{\mathcal{K}}$ defined in Eq.~(\ref{eq:IGMS}), is constructed as follows. Each parameter $\mu_{jk}$ or $\sigma_{jk}$ for $j=1,\ldots,N_\mathrm{INIT}$ and $k=1,\ldots,D_\mathrm{STA}$ is uniformly sampled from its interval $[\mu_{jk}^\mathrm{min}, \mu_{jk}^\mathrm{max}]$ or $[\sigma_{jk}^\mathrm{min}, \sigma_{jk}^\mathrm{max}]$, respectively. The weight vector $\mathbf{w}=[w_1,\ldots,w_{N_\mathrm{INIT}}]^\top$ follows a symmetric Dirichlet distribution, which is simulated by taking $N_\mathrm{INIT}-1$ independent uniform samples from $[0,1]$, including the endpoints 0 and 1, sorting them, and extracting the $N_\mathrm{INIT}$ spacings between consecutive points. 

A transient GMD sample $\mathbb{U}(\widehat{\mathcal{K}}(T_\text{INIT}))$ up to the time horizon $T_\text{INIT}$, as given in Eq.~(\ref{eq:tgms}), is obtained by running Algorithm~\ref{alg:tpaps} of the TPAPS on a sample $\mathbb{U}((0,T_\text{INIT}]\times\mathcal{P}\times\widehat{\mathcal{K}})$. The sample is a triple $(t, \bm{\Theta}, \bm{\Phi}_\text{I})$, where the time $t$ is uniformly sampled in $(0, T_\text{INIT}]$, the system parameters $\bm{\Phi}$ are uniformly sampled in the parameter domain $\mathcal{P}$, and the initial GMD $\bm{\Phi}_\text{I}$ is drawn from the IGMS $\widehat{\mathcal{K}}$.

To sample $n$ GMDs in $\widehat{\mathcal{K}}\cup\widehat{\mathcal{K}}(T_\text{INIT})$ that covers both the IGMS and TGMS, we define the set
\begin{equation}
\mathbb{D}(n, \lambda)= \mathbb{U}(\widehat{\mathcal{K}},\lfloor\lambda \cdot n\rfloor)\cup\mathbb{U}(\widehat{\mathcal{K}}(T_\text{INIT}), n - \lfloor\lambda \cdot n\rfloor),\nonumber
\end{equation}
with the sample fraction $\lambda \in [0,1]$, where $\lfloor\cdot\rfloor$ is the floor function. This set includes $\lfloor\lambda\cdot n\rfloor$ random GMDs in the IGMS and $n-\lfloor\lambda\cdot n\rfloor$ random GMDs in the TGMS. The sample fraction $\lambda$ determines how much the sampler prioritizes producing the initial distributions in the IGMS versus the transient distributions in the TGMS. Since transient distributions may act as initial distributions in recursion, this balance affects long-term prediction fidelity. When $\lambda$ is close to 1, the TPAPS primarily learns initial distributions solely from the IGMS, causing it to only predict within the IGMS up to $t_\text{LEAP}$. Conversely, if $\lambda$ is close to 0, the TPAPS overemphasizes transient dynamics but fails to properly represent the initial distributions in the IGMS, which undermines the recursive foundation and leads to erroneous short-term predictions.

\subsubsection{Loss function}

For a GMD parameterized by $\bm{\Phi}$, the learning goal of the autoencoder is that the input parameters $\bm{\Phi}$ and the restored parameters $\widehat{\bm{\Phi}}=f_\text{D}(f_\text{E}(\bm{\Phi}))$ correspond to the same distribution. The autoencoder loss of this single GMD is defined by the average L1 error between the two PDFs evaluated on $N_\text{AE}$ random states in the state domain $\mathcal{S}$
\begin{equation}
l_\text{AE}(\bm{\Phi}) = \frac{1}{N_\text{AE}}\sum_{\mathbf{x}\in \mathbb{U}(\mathcal{S},N_\text{AE})}\bigg| f_{\text{GMD}}(\mathbf{x};\bm{\Phi}) - f_{\text{GMD}}(\mathbf{x};f_\text{D}(f_\text{E}(\bm{\Phi})))\bigg|.\nonumber
\end{equation}

For any initial or transient GMD $\bm{\Phi}_{\text{I}}\in \widehat{\mathcal{K}}\cup\widehat{\mathcal{K}}(T_\text{INIT})$, system parameters $\bm{\Theta}\in\mathcal{P}$, and time $t\in\widehat{\mathcal{T}}$, the TPAPS $q(\mathbf{x},t;\bm{\Theta},\bm{\Phi}_{\text{I}})$ in Eq.~(\ref{eq:TPAPS_full}) or equivalently, the GMD $\bm{\Phi}_t(\bm{\Theta},\bm{\Phi}_{\text{I}})$, should satisfy the transient FPE in Eq.~(\ref{eq:condition_fpe}) for all states $\mathbf{x}\in\mathcal{S}$. This core target is captured by the average L1 loss of the FPE on $N_\text{FP}$ random states in $\mathcal{S}$, defined by
\begin{equation}
l_\text{FP}(t,\bm{\Theta},\bm{\Phi}_{\text{I}}) = \frac{1}{N_\text{FP}}\sum_{\mathbf{x}\in \mathbb{U}(\mathcal{S},N_\text{FP})}\bigg| \frac{\partial }{\partial t}q(\mathbf{x}, t;\bm{\Theta}, \bm{\Phi}_{\text{I}}) - \mathcal{L}_{\text{FP}} q(\mathbf{x}, t;\bm{\Theta}, \bm{\Phi}_{\text{I}})\bigg|.\nonumber
\end{equation}
As the TPAPS is a continuous function, the derivatives $\partial/\partial t$, $\{\partial/\partial x_i\}_{i=1}^{D_\text{STA}}$ and $\{\partial^2/\partial x_i\partial x_j\}_{i,j=1}^{D_\text{STA}}$ are calculated by automatic differentiation implemented in the PyTorch library.

Besides, we explicitly include a normalization term to avoid degraded learning. This normalization loss is based on the 1-D marginal distributions in Eq.~(\ref{eq:marginal_normalization}), which is an efficient approximation of the $D_\text{STA}$-D normalization condition in Eq.~(\ref{eq:condition_norm}). The normalization loss to constrain the GMD parameters $\bm{\Phi}$ is
\begin{equation}
l_\text{NORM}(\bm{\Phi})=\sum_{k=1}^{D_\text{STA}}\left[\sum_{i=1}^{N_{\text{NORM}}}\left(\sum_{j=1}^{N_\text{GAU}}w_j\mathcal{N}(x_{ik};\mu_{jk},\sigma_{jk}^2)\right)\Delta_k-1\right]^2,\nonumber
\end{equation}
where each of the sets $\{x_{ik}\}_{i}^{N_\text{NORM}}, k=1,\cdots,D_\text{STA}$ defines a uniform grid that covers the interval $[x_k^{\text{min}}, x_k^{\text{max}}]$, the $k$-th side of the state domain $\mathcal{S}$ in Eq.~(\ref{eq:state_domain}), and $\Delta_k=(x_k^{\text{max}} - x_k^{\text{min}}) / N_\text{NORM}$ is the length element.

In each training batch, the loss function of the TPAPS is
\begin{equation}\label{eq:total_loss}
\begin{split}
L_\text{TOTAL} &= \gamma_\text{AE}\cdot L_\text{AE} + L_\text{FP} + L_\text{NORM}\\
&= \gamma_\text{AE}\cdot\left[\frac{1}{B_\text{AE}}\sum_{\bm{\Phi}\in\mathbb{D}(B_\text{AE}, \lambda_\text{AE})}l_\text{AE}(\bm{\Phi})\right] + \frac{1}{B_\text{FP}}\sum_{(t,\bm{\Theta},\bm{\Phi})\in\mathcal{A}}l_\text{FP}(t,\bm{\Theta},\bm{\Phi}) + \frac{1}{B_\text{FP}}\sum_{(t,\bm{\Theta},\bm{\Phi})\in\mathcal{A}}l_\text{NORM}(\bm{\Phi}),\\
\text{where}\quad\mathcal{A}&=\mathbb{U}(\widehat{\mathcal{T}},B_\text{FP})\overline{\parallel}\mathbb{U}(\mathcal{P},B_\text{FP})\overline{\parallel}\mathbb{D}(B_\text{FP},\lambda_\text{FP}).
\end{split}
\end{equation}
In the first term, $L_\text{AE}$ is the average of the autoencoder losses evaluated on $B_\text{AE}$ random GMDs $\bm{\Phi}\in\mathbb{D}(B_\text{AE},\lambda_\text{AE})$ sampled in $\widehat{\mathcal{K}}\cup \widehat{\mathcal{K}}(T_\text{INIT})$, and $\gamma_\text{AE}$ is a penalty coefficient. The second term $L_\text{FP}$ is the average of the FPE losses on $B_\text{FP}$ short-term transient distributions $\bm{\Phi}_t(\bm{\Theta},\bm{\Phi})$ randomly sampled in $\widehat{\mathcal{T}} \times \mathcal{P}\times [\widehat{\mathcal{K}}\cup\widehat{\mathcal{K}}(T_\text{INIT})]$. The third term $L_\text{NORM}$ is the average of the normalization losses evaluated on the $B_\text{FP}$ initial distributions $\bm{\Phi}\in\mathbb{D}(B_\text{FP},\lambda_\text{FP})$ in the second term. The ADAM~\cite{kingma2017adam} algorithm is applied to optimize the TPAPS, i.e., the weights of the four ResNets. The training process is detailed in Algorithm~\ref{alg:training}.

\begin{algorithm*}
\caption{Training the TPAPS}\label{alg:training}
\begin{algorithmic}[1]
\REQUIRE The state domain $\mathcal{S}$, system parameter domain $\mathcal{P}$, the IGMS, time leap $t_\text{LEAP}$, number of training batches $N_\text{BATCH}$.
\FOR {$k=1$ to $N_\text{BATCH}$}
\STATE Sample $B_\text{AE}$ GMDs $\{\bm{\Phi}_i\}_{i=1}^{B_\text{AE}}=\mathbb{D}(B_\text{AE}, \lambda_\text{AE})$ in the IGMS and TGMS.
\STATE Calculate the $B_\text{AE}$ autoencoder losses $\{l_\text{AE}(\bm{\Phi}_i)\}_{i=1}^{B_\text{AE}}$ and their average $L_\text{AE}$.
\STATE Sample $B_\text{FP}$ GMDs $\{\bm{\Phi}_j\}_{j=1}^{B_\text{FP}}=\mathbb{D}(B_\text{FP}, \lambda_\text{FP})$, $B_\text{FP}$ short times $\{t_j\}_{j=1}^{B_\text{FP}}=\mathbb{U}((0,t_\text{LEAP}],B_\text{FP})$, and $B_\text{FP}$ system parameter choices $\{\bm{\Theta}_j\}_{j=1}^{B_\text{FP}}=\mathbb{U}(\mathcal{P},B_\text{FP})$.
\STATE Calculate the $B_\text{FP}$ FPE losses $\{l_\text{FP}(t_j, \bm{\Theta}_j, \bm{\Phi}_j)\}_{j=1}^{B_\text{FP}}$ and their average $L_\text{FP}$.
\STATE Calculate the $B_\text{FP}$ normalization losses $\{l_\text{NORM}(\bm{\Phi}_j)\}_{j=1}^{B_\text{FP}}$ and their average $L_\text{NORM}$.
\STATE Obtain the total loss $\gamma_\text{AE}L_\text{AE}+L_\text{FP}+L_\text{NORM}$.
\STATE Optimize the four ResNets using ADAM algorithm.
\ENDFOR
\RETURN The weights of the EMB-RN, REP-RN, EVO-RN, and DEC-RN.
\end{algorithmic}
\end{algorithm*}

In each batch, the autoencoder is evaluated on $N_\text{AE}B_\text{AE}$ state-distribution pairs and the transient FPEs are evaluated on $N_\text{FP}B_\text{FP}$ state-distribution pairs with distinct times, system parameters and initial distributions in the IGMS and TGMS. Empirically, we set the sample fractions $\lambda_\text{AE}=\lambda_\text{FP}=0.75$ to learn a robust representation of the IGMS while retaining sufficient capacity to encode long-term transient distributions. Thus 25\% of the distributions in $\mathbb{D}(B_\text{AE},\lambda_\text{AE})$ and the initial distributions in $\mathbb{D}(B_\text{FP},\lambda_\text{FP})$ are themselves transient distributions in the TGMS. Under this setting, both the GMD autoencoder and the EVO-RN gain the flexibility to handle widely varying distributions beyond the IGMS, enhancing the generalization ability of the TPAPS. Therefore, the training stage would exhaust the joint domain $\mathcal{S}\times\mathcal{T}\times\mathcal{P}\times\widehat{\mathcal{K}}$. It is worth noting that in Eq.~(\ref{eq:total_loss}), the samples used to train the autoencoder in the first term and those for the EVO-RN in the second term are generated independently. This modularization enables us to improve each component or the sampling strategy independently in future work.

Our preliminary numerical results indicate that among the three loss terms in Eq.~(\ref{eq:total_loss}), the normalization loss $L_\text{NORM}$ rapidly decreases to a very low level during training. In contrast, the autoencoder learns much more slowly. To accelerate the learning of the autoencoder, we choose a large penalty coefficient $\gamma_\text{AE}>1$ and use more samples in training the autoencoder, i.e., $B_\text{AE}>B_\text{FP}$ and $N_\text{AE}>N_\text{FP}$.

\section{Numerical experiments}
\label{sec:4}
This section serves two purposes. First, we investigate a GMD autoencoder without the influence of a specific stochastic system. Subsequently, we deploy the full TPAPS framework on three paradigmatic stochastic systems to learn their complete dynamical portrait, encompassing both transient and stationary behaviors under varying initial conditions and system parameters. The proposed TPAPS $q$ is compared with MCS $p$ by using the Euler-Maruyama method, which is detailed in Appendix~\ref{sec:euler_maruyama}. The L1 error
\begin{equation}
\int_\mathcal{S}|q(\mathbf{x},t;\mathbf{\Theta},\mathbf{\Phi}_\text{I})-p(\mathbf{x},t;\mathbf{\Theta},\mathbf{\Phi}_\text{I})|\mathrm{d}\mathbf{x},
\end{equation}
between distributions $p$ and $q$ is utilized as the quantitative metric.

\subsection{The GMD autoencoder without dynamics}
\label{sec:visul_gmd_ae}
\begin{figure}[!htb]
\center{\includegraphics[width=1\textwidth]
{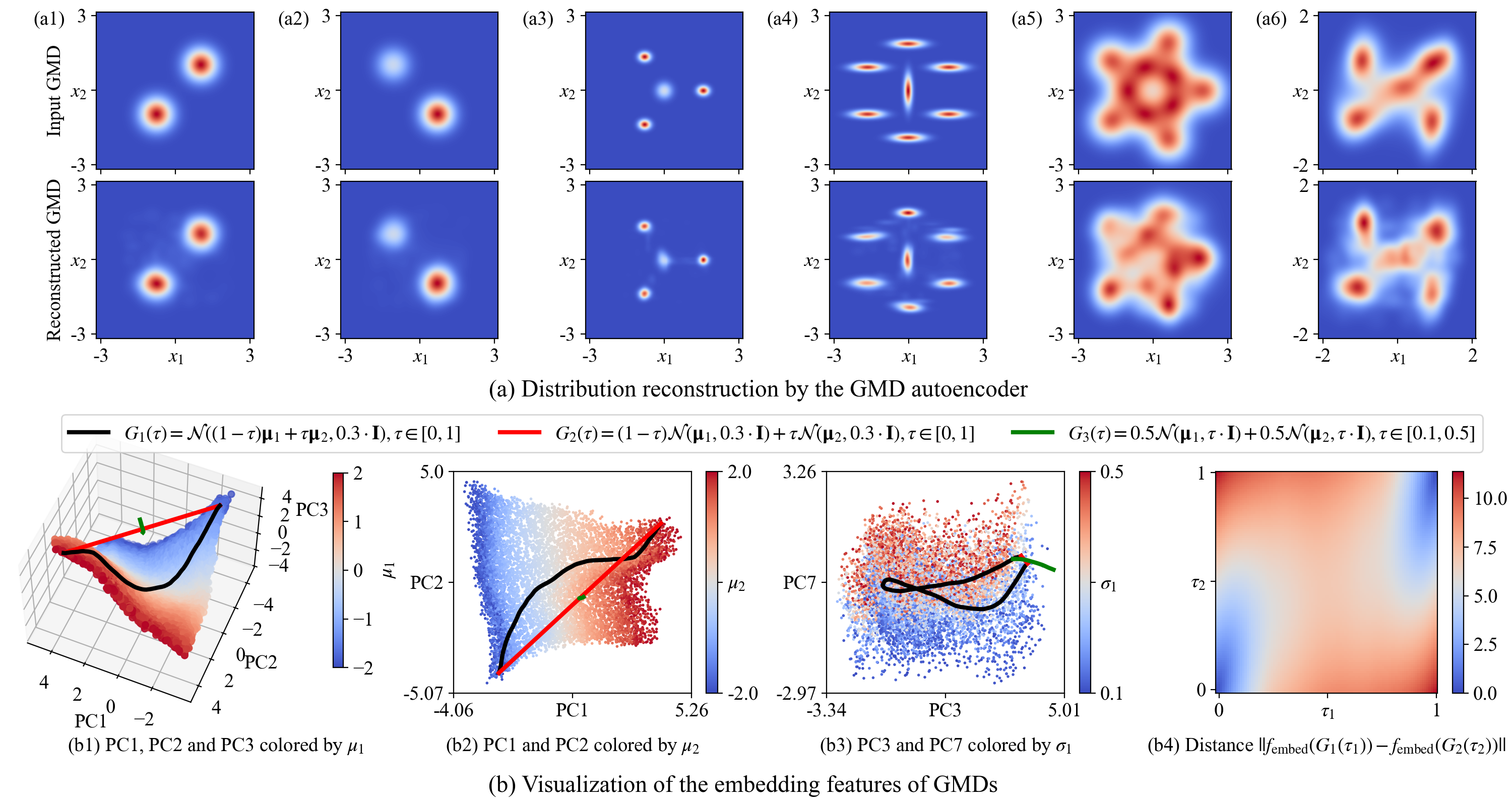}}
\caption{\label{fig:gmd_encoder} Visualization of the GMD autoencoder. (a) Six GMDs and their reconstructions. (b) The PCA and distance visualizations of the 50-D embedding space. Each point represents a 2-D Gaussian distribution in Eq.~(\ref{eq:IGMS_pure_ae}). The three lines consist of continuously changing distributions, where $\bm{\mu}_1=[-2, -2]^\top$, $\bm{\mu}_2=[2, 2]^\top$, and $\mathbf{I}$ is the 2-D identity matrix.}
\end{figure}

Before integrating the GMD autoencoder into the framework of solving FPEs, we train a 2-D GMD autoencoder without considering stochastic dynamics to show its effectiveness and limitations in modeling complex distributions. The GMDs for training are randomly chosen in the IGMS $\widehat{\mathcal{K}}$
\begin{equation}\label{eq:IGMS_pure_ae}
\widehat{\mathcal{K}}=\{\mathbf{\Phi}|[w_1,\cdots,w_5]^\top\in \mathcal{C}(5), \mu_{jk}\in [-2, 2], \sigma_{jk}\in[0.1, 0.5],j=1,\cdots,5,k=1,2\},
\end{equation}
consisting of GMDs with $N_\text{INIT}=5$ components where the means are in the interval $[-2,2]$ and the standard deviations in $[0.1, 0.5]$. The loss function only consists of the autoencoder loss and the normalization loss
\begin{equation}\label{eq:loss_pure_ae}
L_\text{AE} + L_\text{NORM} = \frac{1}{B_\text{AE}}\sum_{\bm{\Phi}\in\mathbb{U}(\widehat{\mathcal{K}}, B_\text{AE})}[l_\text{AE}(\bm{\Phi})+l_\text{NORM}(\bm{\Phi})].
\end{equation}
The normalization loss is calculated on a grid of $50^2=2500$ states covering the square state domain $[-5, 5]\times [-5, 5]$. The ADAM algorithm is utilized to optimize the autoencoder with the learning rate 0.0002 and the batch size $B_\text{AE}=700$ for $10^6$ batches. The reconstruction results are shown in Fig.~\ref{fig:gmd_encoder}.

Figure~\ref{fig:gmd_encoder}(a) demonstrates 6 GMDs and their reconstructions by the autoencoder. Figures~\ref{fig:gmd_encoder}(a1)-(a3) show that GMDs with 2 or 4 components with various weights and SDs can be well reconstructed. Impressively, Fig.~\ref{fig:gmd_encoder}(a4) indicates that the autoencoder can be applied to GMDs with 7 components, while the training set only consists of GMDs with 5 components. Due to the convexity-preserving embedding in Eq.~(\ref{eq:condition_convex_1}), the autoencoder can effectively generalize to more complex distributions, such as the 10-component GMD in Fig.~\ref{fig:gmd_encoder}(a5) and the transient GMD in Fig.~\ref{fig:gmd_encoder}(a6). However, the reconstruction qualities are considerably poor, indicating that for unseen distributions with many Gaussian components, generalization is limited when they differ significantly from the training set. This observation suggests that the training set of the autoencoder should also include the plausible transient distributions of the stochastic systems, motivating our autoencoder loss $L_\text{AE}$ in Eq.~(\ref{eq:total_loss}), which considers both the IGMS $\widehat{\mathcal{K}}$ and the TGMS $\widehat{\mathcal{K}}(T_\text{INIT})$.

In Fig.~\ref{fig:gmd_encoder}(b), the $50$-D embedding space of the GMD autoencoder is visualized by processing 20,000 Gaussian distributions $\mathcal{N}([\mu_1,\mu_2]^\top,\text{diag}\{\sigma_1^2,\sigma_2^2\})$ with random means $\mu_1,\mu_2\in[-2, 2]$ and SDs $\sigma_1,\sigma_2\in[0.1, 0.5]$, then applying the principal component analysis (PCA), and finally plotting the first several principal components (PCs) of the embedding vectors $\mathbf{h}$. Figure~\ref{fig:gmd_encoder}(b1) shows that the embeddings of these individual Gaussians form the boundary of the convex hull, whose interior corresponds to the multi-component Gaussians. The coloring in Figs.~\ref{fig:gmd_encoder}(b1) and (b2) indicates that the leading 1st and 2nd PCs represent the means $\mu_1$ and $\mu_2$ of the Gaussian distributions. In contrast, the SDs $\sigma_1$ and $\sigma_2$ that describe the randomness are mainly reflected in the later PCs with smaller variance, as shown in Fig~\ref{fig:gmd_encoder}(b3).

Figure~\ref{fig:gmd_encoder}(b) also details the embeddings of three groups of continuously changing distributions, defined as follows
\begin{equation}\label{eq:three_lines}
\begin{split}
G_1(\tau)&=\mathcal{N}((1-\tau)\cdot\bm{\mu}_1+\tau\cdot\bm{\mu}_2,0.3\cdot\mathbf{I}), \tau\in[0,1],\\
G_2(\tau)&=(1-\tau)\cdot\mathcal{N}(\bm{\mu}_1,0.3\cdot\mathbf{I})+\tau\cdot\mathcal{N}(\bm{\mu}_2,0.3\cdot\mathbf{I}), \tau\in[0,1],\\
G_3(\tau)&=0.5\cdot\mathcal{N}(\bm{\mu}_1,\tau\cdot\mathbf{I})+0.5\cdot\mathcal{N}(\bm{\mu}_2,\tau\cdot\mathbf{I}), \tau\in[0.1,0.5],
\end{split}
\end{equation}
where $\mathbf{I}=\text{diag}\{1, 1\}$ is the identity matrix. The group $G_1$ includes individual Gaussian distributions with the mean vectors on the line segment between $\bm{\mu}_1$ and $\bm{\mu}_2$. Their embeddings are on the boundary of the convex hull in Fig.~\ref{fig:gmd_encoder}(b1). The group $G_2$ consists of 2-component GMDs with various combination weights. Their embeddings are precisely on the line segment that connects the embeddings of the boundary distributions $G_2(0)$ and $G_2(1)$, and the center of the line segment is exactly the embedding of the middle distribution $G_2(0.5)=G_3(0.3)$. This demonstrates clearly the core property of the GMD autoencoder in Eq.~(\ref{eq:condition_convex_1}), i.e., the embedding and the convex combination commute. Consequently, the combination proportion $\tau$ is preserved as a linear coordinate along the line segment in the embedding space. Figure~\ref{fig:gmd_encoder}(b4) further illustrates the validity of the embedding: similar distributions are close to each other in the embedding space. When $\tau$ approaches 0 or 1, both $G_1(\tau)$ and $G_2(\tau)$ converge to the same Gaussian distribution. Consequently, the L2 distance between their embedding features approaches zero. 

The distributions in $G_3$ have fixed means and variable SDs. Figure~\ref{fig:gmd_encoder}(b) shows that their embedding vectors vary only slightly and remain near the point $f_\text{embed}(G_3(0.3))=f_\text{embed}(G_2(0.5))$. This is consistent with the earlier observation that the leading PCs capture the locality of the GMD, while the later components encode variance information. Because the means are fixed in $G_3$, the embeddings move only in the directions of the later PCs, resulting in small changes.

A distinct feature of the embedding space is that embeddings of GMDs with many components tend to lie near the origin, while individual Gaussians form the convex hull. This can be understood recursively. The embedding of an $n$-component GMD can be obtained through $n-1$ convex combinations of embeddings of simpler GMDs. Because the convex combination satisfies
\begin{equation}\label{eq:shrink}
\|(1-\tau)\mathbf{a}+\tau\mathbf{b}\|\leq (1-\tau)\|\mathbf{a}\| + \tau\|\mathbf{b}\|\leq \max\{\|\mathbf{a}\|,\|\mathbf{b}\|\},
\end{equation}
for any vectors $\mathbf{a}$ and $\mathbf{b}$ and the weights $\tau\in(0,1)$, the norm of the resulting embedding is generally no greater than the larger norm of the two vectors being combined. If the two vectors have equal norm and are not aligned, the norm of the combination is strictly smaller, explaining the contraction toward the origin. Repeating this process for many components typically drives the embedding toward the origin, explaining why multi-component GMDs cluster near the center.

In the following, the training set of the GMD autoencoder will include not only the IGMS but also the TGMS, which would significantly adapt and improve its representation capability.

\subsection{1-D system}
\label{sec:system_1d}
We consider the 1-D system~\cite{Wang2025Pseudo}
\begin{equation}\label{eq:sys_1d}
\dot{x}=ax^5+bx^4+cx^3+dx^2+ex+f + \sigma\xi, \quad a<0,
\end{equation}
with seven system parameters $\bm{\Theta}=[a, b, c, d, e, f, \sigma]^\top$, where $\xi$ is a SWGN and $\sigma$ is the noise amplitude. The corresponding FP operator of the transient solution $p=p(x,t;\bm{\Theta},p_\text{I})$ is
\begin{equation}
\mathcal{L}_\text{FP}p=-\frac{\partial}{\partial x}[(ax^5+bx^4+cx^3+dx^2+ex+f)p]+\frac{\sigma^2}{2}\frac{\partial^2p}{\partial x^2}.\nonumber
\end{equation}
By setting the potential function
\begin{equation}
U(x;\bm{\Theta})=-\frac{ax^6}{6}-\frac{bx^5}{5}-\frac{cx^4}{4}-\frac{dx^3}{3}-\frac{ex^2}{2}-fx,\nonumber
\end{equation}
the system~(\ref{eq:sys_1d}) has the analytical stationary distribution
\begin{equation}\label{eq:spdf_1d_sys}
p_\text{S}(x;\bm{\Theta})= C\cdot\text{exp}\left\{-\frac{2U(x;\bm{\Theta})}{\sigma^2}\right\}.
\end{equation}

We choose a large 1-D state domain $\mathcal{S}=\{x|x\in[-6,6]\}$ to enclose the significant probability mass and the 7-D parameter domain $\mathcal{P}=\{a\in[-2.5, -0.5];b,c,d,e,f\in[-1, 1];\sigma\in[0.2,2.2]\}$, ranging a broad variety of transient dynamics. The IGMS $\widehat{\mathcal{K}}$ in Eq.~(\ref{eq:IGMS}) includes GMDs with $N_\text{INIT}=5$ components with means $\mu_{j}\in[-3, 3]$ and standard deviations $\sigma_j\in[0.1,0.3]$ for $j=1,\cdots,5$. A TPAPS is trained for $10^6$ batches and the sizes of samples in each batch are $B_\text{FP}=150$, $B_\text{AE}=750$, $N_\text{FP}=150$, $N_\text{AE}=750$, $N_\text{NORM}=200$. The penalty coefficient $\gamma_\text{AE}=5$. The learning rate of the ADAM optimizer is 0.0002. The short-term leap time is $t_\text{LEAP}=1$ and the TGMS includes transient GMDs with the maximal initial time $T_\text{INIT}=3$. Therefore, the expected time horizon for accurate prediction is $T_\text{max}=T_\text{INIT}+t_\text{LEAP}=4$.

\begin{figure}[!htb]
\center{\includegraphics[width=1\textwidth]
{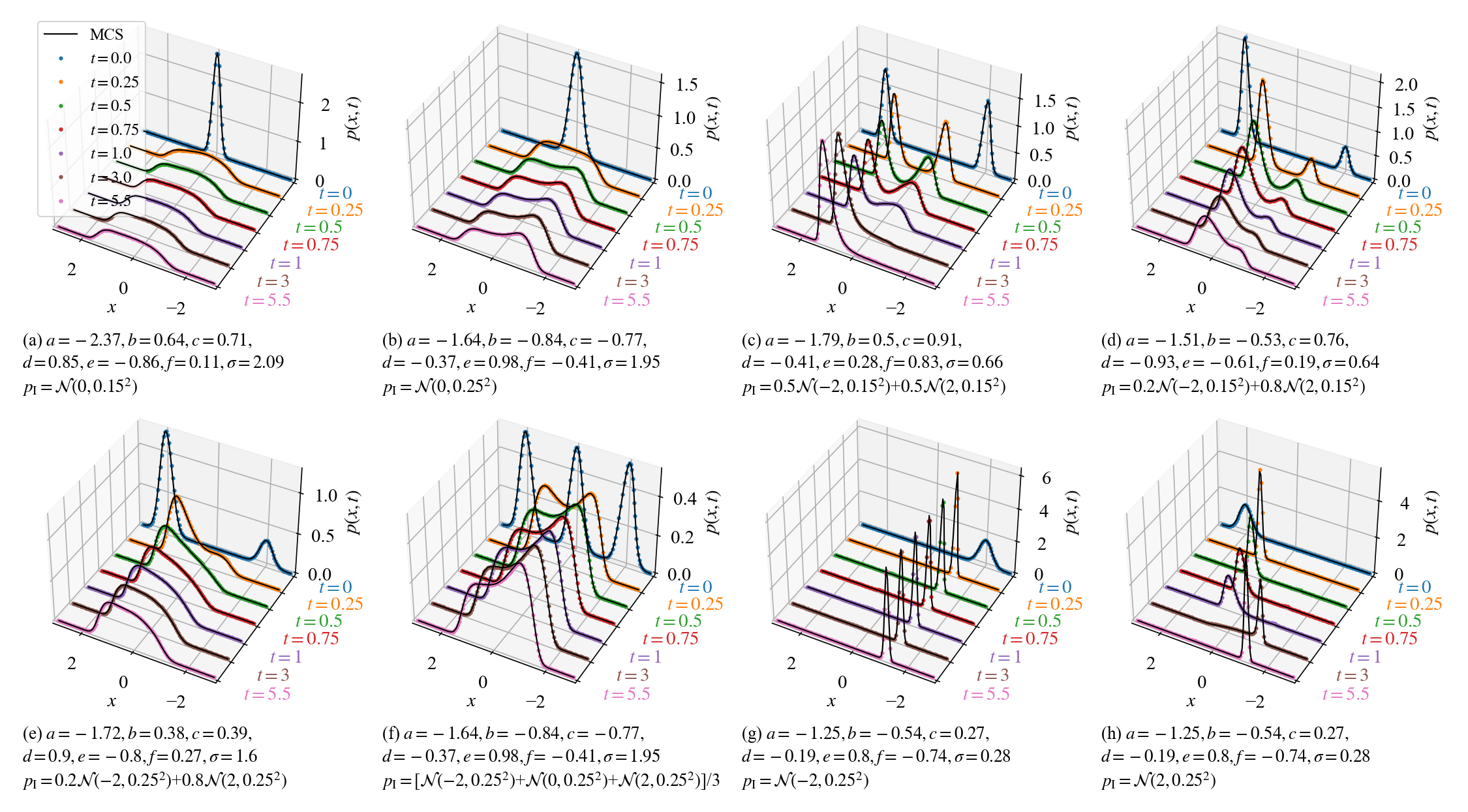}}
\caption{\label{fig:1d_sys_examples} The transient solutions of the 1-D system~(\ref{eq:sys_1d}) by TPAPS (colored circles) and MCS (black lines) with different initial distributions and system parameters.}
\end{figure}

Figure~\ref{fig:1d_sys_examples} details eight groups of transient distributions calculated by the TPAPS with different system parameters and initial distributions, which include 1 to 3 Gaussian components. The true distributions are estimated by the MCS over $10^7$ independent trajectories, as detailed in~\ref{sec:euler_maruyama}.

In all the eight cases, the transient solutions at all different times by the TPAPS show strong agreement with the MCS results, indicating its effectiveness. Even at $t=5.5$, which is significantly beyond $T_\text{max}=4$, TPAPS demonstrates its capability to provide accurate long-term predictions. This is because the systems approach their stationary distributions for $t>4$, beyond which the transient solutions remain stable over time. The dynamics before $t<4$, which were captured during training, are sufficient for our recursive calculation to extrapolate these steady behaviors effectively.

Notably, Figs.~\ref{fig:1d_sys_examples}(g) and (h) show two cases with identical system parameters but different initial distributions $\mathcal{N}(-2, 0.25^2)$ and $\mathcal{N}(2, 0.25^2)$, respectively. The transient solutions generated by TPAPS are both observed to correctly converge to the same unique stationary distribution in Eq.~(\ref{eq:spdf_1d_sys}), which demonstrates its capability in capturing the ergodic nature of the underlying process.

\begin{figure}[!htb]
\center{\includegraphics[width=1\textwidth]
{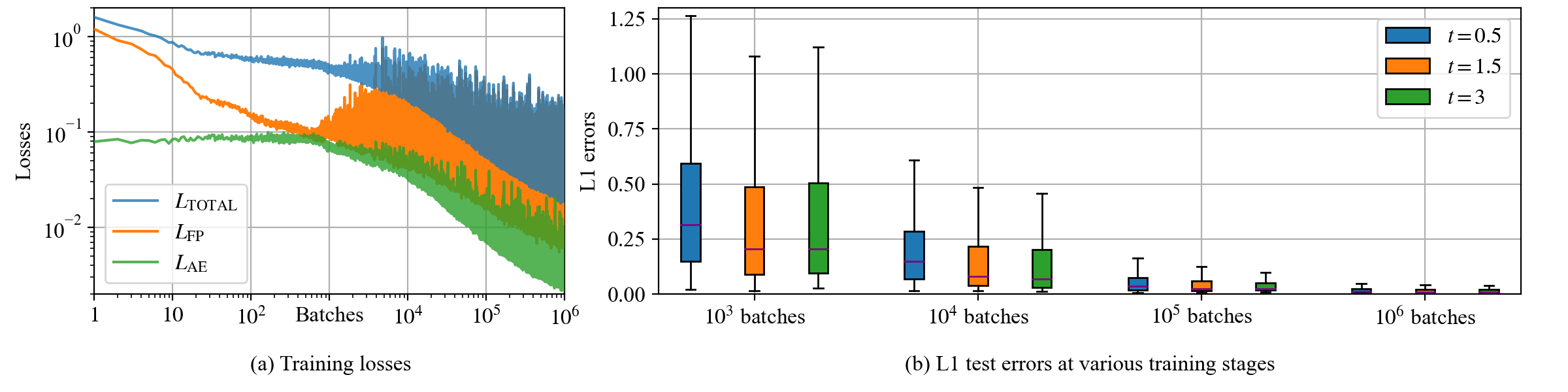}}
\caption{\label{fig:1d_sys_loss} The training losses and test errors of the TPAPS on the 1-D system. The boxplots in (b) are summarized on 2,500 transient solutions with random initial distributions and system parameters.}
\end{figure}

Figure~\ref{fig:1d_sys_loss} demonstrates the stable training process of the TPAPS. Figure~\ref{fig:1d_sys_loss}(a) shows the total loss and the losses $L_\text{FP}$ and $L_\text{AE}$ across the training stage. All loss curves exhibit a steady, albeit fluctuating, downward trend. This behavior stems from the inherent sparsity of samples within each training batch relative to the vast and complex transient dynamics of the system under diverse parameters and initial distributions. To quantitatively evaluate the TPAPS, we randomly sample 2,500 initial distributions in the IGMS $\widehat{\mathcal{K}}$ and 2,500 system parameters in $\mathcal{P}$, and calculate the L1 distances between the distributions of the TPAPS and MCS. Figure~\ref{fig:1d_sys_loss}(b) shows the boxplots of the errors at $t=0.5,1.5$ and 3, and the means, SDs and medians of these extensive tests are summarized in Tab.~\ref{tab:error_1D}. A tenfold increase in training batches from $10^4$ to $10^5$ reduces the mean L1 error to one-third of its prior value. A subsequent tenfold increase to $10^6$ yields a similar relative reduction. This pattern of diminishing returns suggests that the accuracy improvements of the TPAPS are steadily slowing toward a limit.

\begin{table}[!htb]
\footnotesize
\centering
\caption{The mean, SD, and median (MEAN/SD/MED) of the L1 errors between the TPAPS and MCS results across 2,500 random transient solutions of the 1-D system~(\ref{eq:sys_1d}) at various training stages.}\label{tab:error_1D}
\begin{tabular}{ccccc}
\toprule
Time & $10^3$ batches & $10^4$ batches & $10^5$ batches & $10^6$ batches\\
\midrule
$t=0.5$ & 0.4257/0.3561/0.3161 & 0.2062/0.1906/0.1476 & 0.0595/0.0923/0.0345 & 0.0212/0.0477/0.0133\\
$t=1.5$ & 0.3504/0.3701/0.2056 & 0.1794/0.2354/0.0798 & 0.0607/0.1291/0.0232 & 0.0217/0.0645/0.0102\\
$t=3.0$ & 0.3637/0.3745/0.2050 & 0.1648/0.2340/0.0685 & 0.0619/0.1410/0.0230 & 0.0234/0.0717/0.0098\\
\bottomrule
\end{tabular}
\end{table}

The training speed of the TPAPS is 0.82 seconds per batch on a system equipped with an Intel i9-13900K CPU and an NVIDIA GeForce RTX 4090 GPU and $10^5$ batches required 23 hours. While training is relatively time-consuming, the evaluation process is extremely efficient. Specifically, calculating 2,500 transient solutions with random initial distributions and control parameters, where each solution includes snapshots on a grid of 200 states at 12 distinct time points within $[0,3]$, takes only 1.37 seconds with the TPAPS. Within this time, 1.2 seconds are devoted to recursively computing the $2500\times 12 = 30,000$ GMDs, and the remaining 0.17 seconds are used to evaluate the corresponding probability densities over the state grid by Eq.~(\ref{eq:gaussian_mixture}). In comparison, the MCS with $10^7$ trajectories requires approximately 993 seconds on the CPU or 20.4 seconds on the GPU to compute a single transient solution under the same conditions. Thus, the TPAPS achieves a speedup of several thousand times compared to the GPU-accelerated MCS, and is about six orders of magnitude faster than its CPU-based counterpart.

\begin{figure}[!htb]
\center{\includegraphics[width=1\textwidth]
{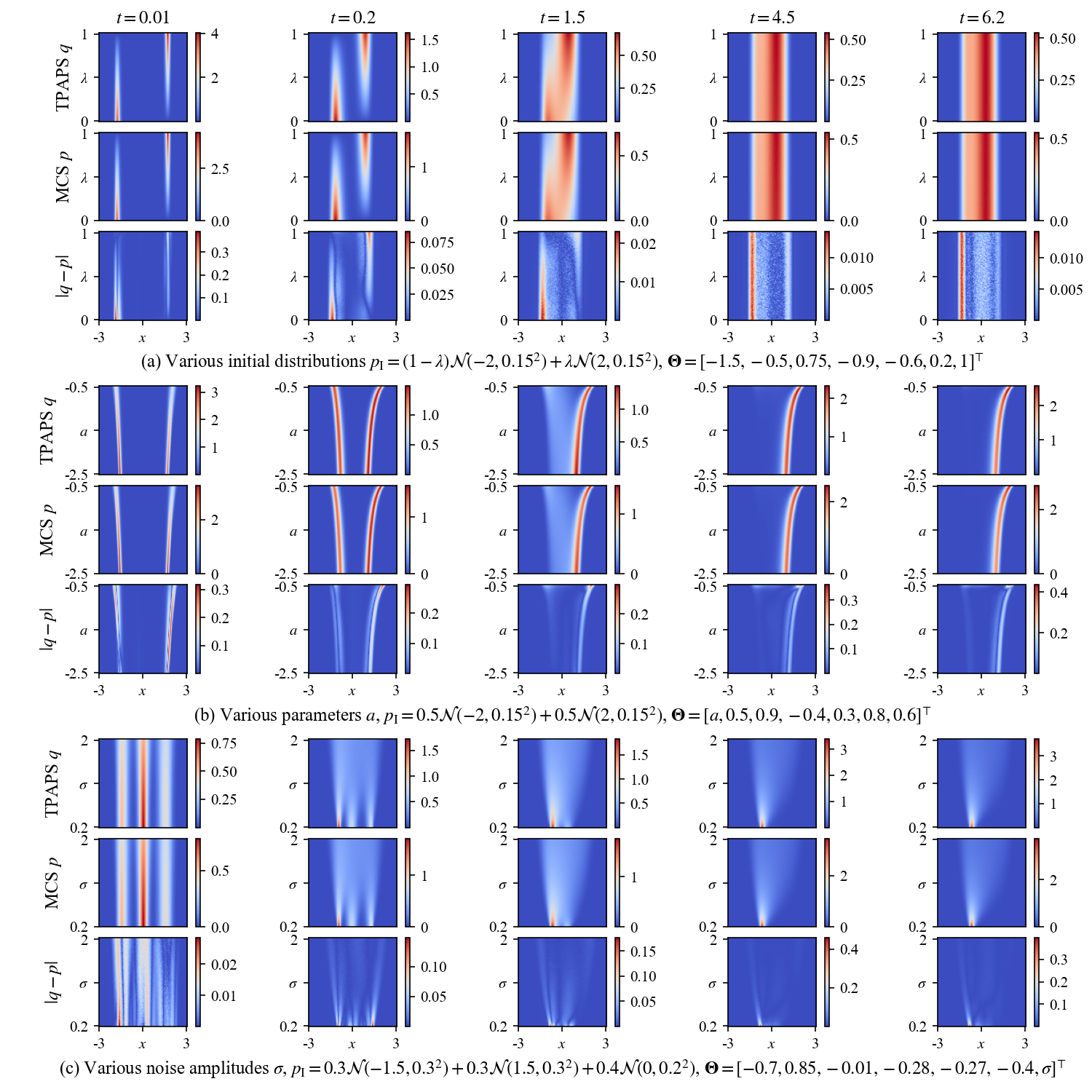}}
\caption{\label{fig:1d_sys_bifu}Bifurcation study of the 1-D system~(\ref{eq:sys_1d}) by using TPAPS and MCS. The bifurcation parameters are (a) the combination weight $\lambda$ in the initial distribution, (b) the system parameter $a$, and (c) the noise amplitude $\sigma$. The resolution of each figure is $100\times 100$.}
\end{figure}

A key advantage of the TPAPS is its ability to rapidly generate a large ensemble of transient solutions, enabling comprehensive exploration of stochastic dynamics across diverse initial conditions and control parameters. Figure~\ref{fig:1d_sys_bifu} shows three examples. In each case, one factor of the system changes with 100 different values. The corresponding distributions at 100 states equally distributed in $[-3, 3]$ are calculated at the times $t=0.01$, 0.2, 1.4, 4.5, and 6.2. In Fig.~\ref{fig:1d_sys_bifu}(a), the initial distributions switch smoothly between the two Gaussian distributions $\mathcal{N}(-2, 0.25)$ and $\mathcal{N}(2, 0.25)$ by tuning the combination weight $\lambda$. When $t$ grows, all the transient distributions converge to the same stationary distribution. In Fig.~\ref{fig:1d_sys_bifu}(b), the parameter $a$ changes, affecting the peak location of the stationary distribution. In Fig.~\ref{fig:1d_sys_bifu}(c), the increase of the noise amplitude $\sigma$ increases the variance of the stationary distribution. In all cases, the TPAPS results agree with the MCS solutions, indicating that the TPAPS captures diverse stochastic dynamics with reasonable accuracy. Note that even on a CPU, the TPAPS requires only about 0.34 seconds to compute the 100 distributions in each of the three cases. Thus, once trained, the TPAPS can be deployed efficiently without GPU acceleration.

\subsection{2-D subcritical system}
\label{sec:system_subcritical}
We next consider a 2-D system~\cite{Attar2009Direct}, which shows the subcritical bifurcation. In the polar coordinates $(r, \theta)$, the system equations are
\begin{equation}\label{eq:subcri_sys_polar}
\dot{r}=\mu r + r^3 - r^5, \quad \dot{\theta}=\omega+b r^2,
\end{equation}
where the parameter $\omega$ governs the oscillation frequency, while $b$ controls the nonlinearity of the oscillation amplitude. When $\mu < -1/4$, the system exhibits a stable attractor at the origin ($r=0$). For $-1/4 < \mu < 0$, in addition to the stable attractor at the origin, there exists a stable limit cycle oscillation (LCO) with radius $r = \sqrt{(1 + \sqrt{1 + 4\mu})/2}$. Once $\mu > 0$, only this stable LCO remains. Thus, the deterministic system~(\ref{eq:subcri_sys_polar}) undergoes a subcritical Hopf bifurcation at $\mu = 0$.

We represent the system~(\ref{eq:subcri_sys_polar}) in the Cartesian coordinates $\mathbf{x}=[x, y]^\top$ and add Gaussian white noises to the bifurcation parameter $\mu$, as well as the Cartesian velocities of $x$ and $y$. Then, the Stratonovich-type SDEs are
\begin{equation}
\begin{split}
\dot{x}(t)&=\left[\mu+\sqrt{2D_\mu}\circ\xi_{\mu}(t)\right] x+(1-r^2)r^2x -\omega y - br^2y+\sqrt{2D}\circ\xi_x(t),\\
\dot{y}(t)&=\left[\mu+\sqrt{2D_\mu}\circ\xi_{\mu}(t)\right] y+(1-r^2)r^2y +\omega x + br^2x+\sqrt{2D}\circ\xi_y(t),
\end{split}\nonumber
\end{equation}
where $r=\sqrt{x^2+y^2}$ and $\xi_\mu$, $\xi_x$, and $\xi_y$ are SWGNs. The noise intensity on $\mu$ is $D_\mu$ and the noise intensities on the velocities are $D$. The corresponding Ito-type SDEs with the Wong-Zakai correction are~\cite{Attar2009Direct}
\begin{equation}
\begin{split}\label{eq:subcri_sys}
\dot{x}(t)&=\mu x +(1-r^2)r^2x -\omega y - br^2y+D_\mu x+\sqrt{2D_\mu}x\xi_{\mu}(t)+\sqrt{2D}\xi_x(t),\\
\dot{y}(t)&=\mu y +(1-r^2)r^2y +\omega x + br^2x+D_\mu y+\sqrt{2D_\mu}y\xi_{\mu}(t)+\sqrt{2D}\xi_y(t),
\end{split}
\end{equation}
with five parameters $\bm{\Theta}=[\mu, b, \omega, D, D_\mu]^\top$. The FP operator of the transient distribution $p=p(\mathbf{x},t;\bm{\Theta}, p_\text{I})$ is
\begin{equation}
\begin{split}
\mathcal{L}_\text{FP}p&=-\frac{\partial}{\partial x}\left\{\left[\mu x +(1-r^2)r^2x -\omega y - br^2y+D_\mu x\right]p\right\}\\
&- \frac{\partial}{\partial y}\left\{\left[\mu y +(1-r^2)r^2y +\omega x + br^2x+D_\mu y\right]p\right\}\\
&+\frac{\partial^2}{\partial x^2}\left[(D+D_\mu x^2)p\right]
+ \frac{\partial^2}{\partial x\partial y}\left[(2D_\mu x y) p\right]
+\frac{\partial^2}{\partial y^2}\left[(D+D_\mu y^2)p\right].
\end{split}\nonumber
\end{equation}

We choose the 2-D state domain $\mathcal{S}=\{\mathbf{x}=[x, y]^\top|x,y\in[-3,3]\}$ and the 5-D parameter domain $\mathcal{P}=\{\mu\in[-0.4, 0.4],b\in[0,0.02], \omega\in[0.5, 1.5], D\in[0.01, 0.3], D_\mu\in[0, 0.05]\}$. The IGMS $\widehat{\mathcal{K}}$ in Eq.~(\ref{eq:IGMS}) includes GMDs with $N_\text{INIT}=5$ 2-D Gaussian components with means $\mu_{jk}\in[-1, 1]$ and standard deviations $\sigma_{jk}\in[0.1,0.3]$ for $j=1,\cdots,5$ and $k=1,2$. A TPAPS is trained for $10^6$ batches and the sizes of samples in each batch are $B_\text{FP}=150$, $B_\text{AE}=750$, $N_\text{FP}=150$, $N_\text{AE}=750$, $N_\text{NORM}=200$. The leap time is $t_\text{LEAP}=1$ and the maximal initial time is $T_\text{INIT}=6$. The expected time horizon for accurate prediction is $T_\text{max}=T_\text{INIT}+t_\text{LEAP}=7$. The penalty coefficient $\gamma_\text{AE}=5$ and the learning rate of the ADAM optimizer is 0.001.

\begin{figure}[!htb]
\center{\includegraphics[width=1\textwidth]
{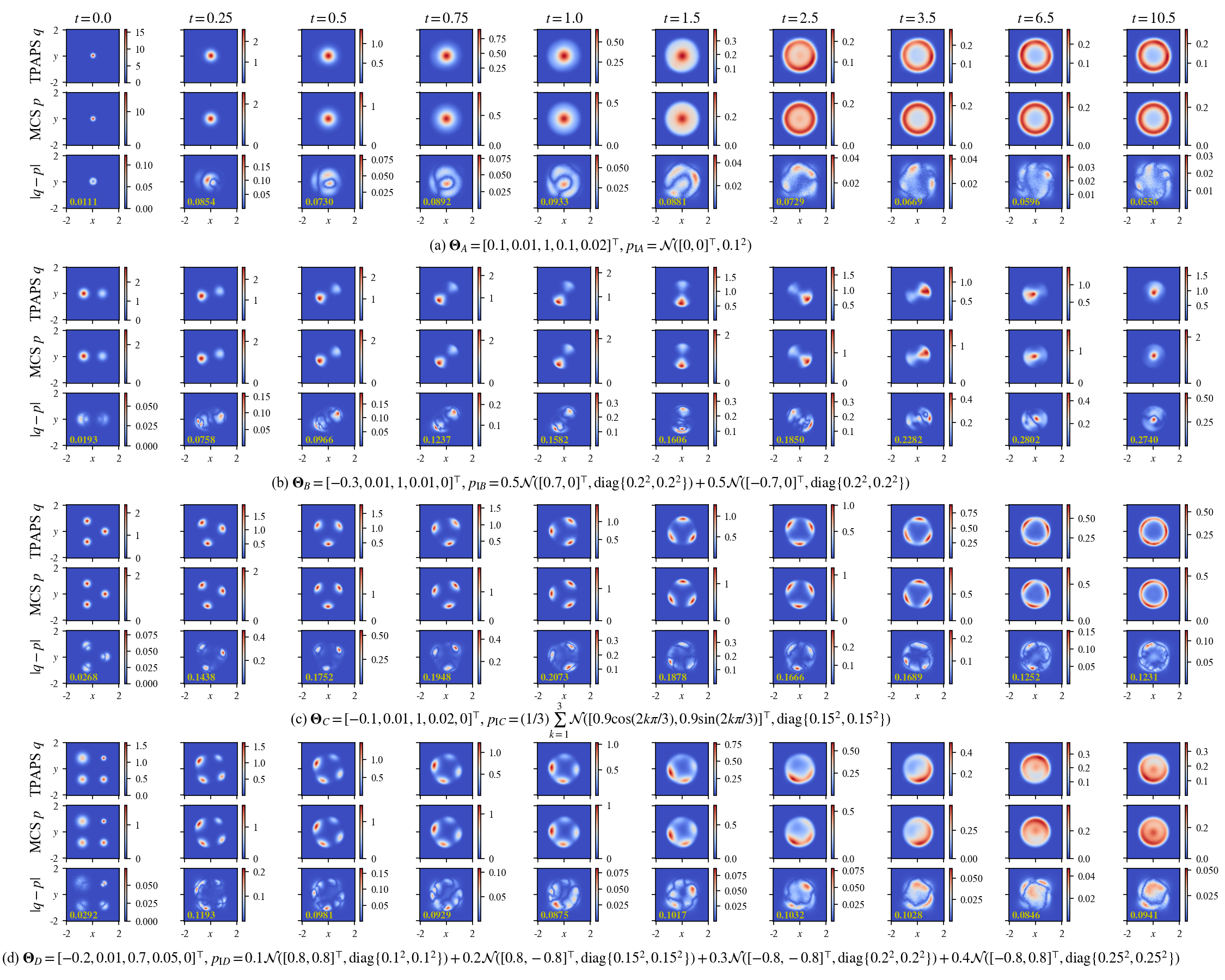}}
\caption{\label{fig:2d_subcri_sys_examples}Four transient solutions of the 2-D subcritical system~(\ref{eq:subcri_sys}) by TPAPS and MCS with different initial distributions and system parameters. The L1 errors are displayed on the residual images.}
\end{figure}

Figure~\ref{fig:2d_subcri_sys_examples} compares four groups of transient solutions calculated by the TPAPS and the MCS up to $t=10.5$. The four challenging initial distributions $p_{\text{I}A}$, $p_{\text{I}B}$, $p_{\text{I}C}$ and $p_{\text{I}D}$ include 1 to 4 Gaussian components with different weights, respectively. When $t=0$, the TPAPS only calls the autoencoder to reconstruct the initial distributions with high accuracy, as can be seen in the first column of Fig.~\ref{fig:2d_subcri_sys_examples}. When $t>0$, the TPAPS correctly recovers the respective rotating transient distributions with multiple components and varying phase angles in all cases. When $t>10$, the three cases (a), (b), and (d) almost converge to the stationary distributions, denoted by $p_{\text{S}A}$, $p_{\text{S}B}$, and $p_{\text{S}D}$, respectively, while the case (c) exhibits persistent oscillations. The TPAPS correctly converges to these stationary distributions or oscillations, demonstrating its capability to resolve long-term dynamics.

A key feature of the TPAPS is that diverse initial and transient distributions are embedded into the same unconstrained vectorial embedding space by the EMB-RN and the refined representation space by the REP-RN. This key achievement is because the GMD autoencoder of the TPAPS is trained on both the IGMS and the TGMS, in contrast to Fig.~\ref{fig:gmd_encoder}, where the autoencoder was trained solely on the IGMS. The huge variety of training distributions in the TGMS significantly expand the capacity of the embedding space. 

\begin{figure}[!htb]
\center{\includegraphics[width=1\textwidth]
{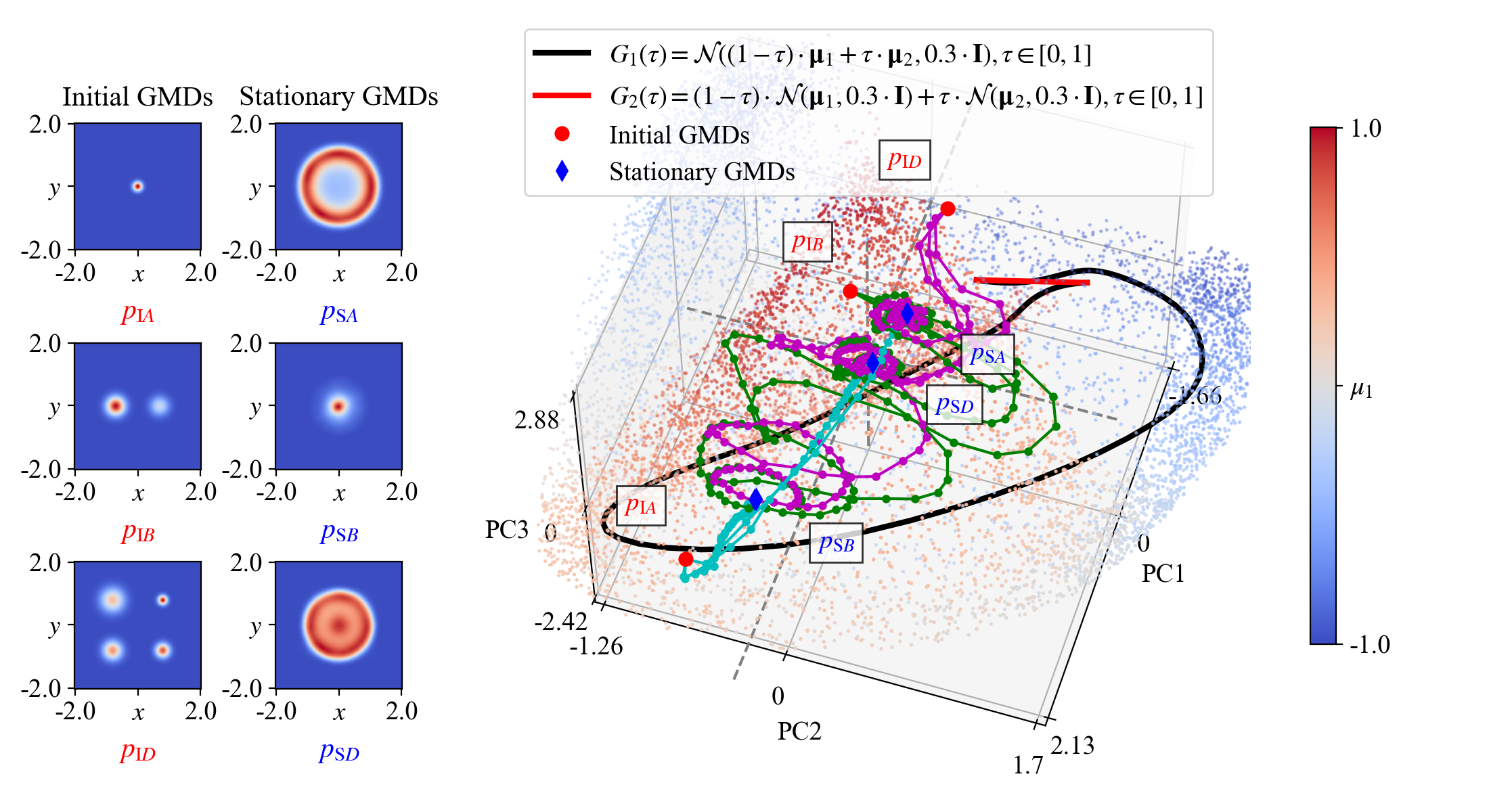}}
\caption{\label{fig:2d_subcritical_sys_embedding}Visualization of transient solutions of the 2-D subcritical system~(\ref{eq:subcri_sys}) with $t\in[0,20]$ in the 1st, 2nd, and 3rd PCs of the 50-D embedding space. The 9 transient solutions initialize from the distributions $p_{\text{I}A}$, $p_{\text{I}B}$, and $p_{\text{I}D}$ in Fig.~\ref{fig:2d_subcri_sys_examples} with system parameters $\bm{\Theta}_A$, $\bm{\Theta}_B$, and $\bm{\Theta}_D$ and converge to three stationary distributions $p_{\text{S}A}$, $p_{\text{S}B}$, and $p_{\text{S}D}$, which are attaching to the respective system parameters. For better illustration, similar to Fig.~\ref{fig:gmd_encoder}, the points in the background are embedding features of random Gaussian distributions $\mathcal{N}([\mu_1, \mu_2]^\top,\text{diag}\{\sigma_1^2,\sigma_2^2\})$, and the two lines of distributions $G_1$ and $G_2$ are defined in Eq.~(\ref{eq:three_lines}). The grey dashed lines indicate the axes of the three leading PCs.}
\end{figure}

By using the PCA, Fig.~\ref{fig:2d_subcritical_sys_embedding} visualizes the first three PCs of the 50-dimensional TPAPS embeddings of the 2-D subcritical system~(\ref{eq:subcri_sys}). The colorful points in the background represent embeddings of one-component Gaussian distributions with random means in $[-2, 2]$ and random standard deviations in $[0.1, 0.5]$, consistent with the setup in Fig.~\ref{fig:gmd_encoder}. Compared with Fig.~\ref{fig:gmd_encoder}(b1), however, the distributions in Fig.~\ref{fig:2d_subcritical_sys_embedding} are scattered within a much more complex convex hull. The central region of the embedding space is now filled by practical transient GMDs with multiple components, rather than remaining a hollow void as in Fig.~\ref{fig:gmd_encoder}(b1).

To better visualize the evolution of transient solutions in the embedding space, Fig.~\ref{fig:2d_subcritical_sys_embedding} plots the embedded trajectories of 9 transient solutions generated by the TPAPS. Each trajectory has 81 points, corresponding to the distributions at the times $t\in[0, 20]$ with a temporal increment 0.25. These solutions originate from three initial GMDs $p_{\text{I}A}$, $p_{\text{I}B}$, and $p_{\text{I}D}$, each with system parameters $\bm{\Theta}_{A}$, $\bm{\Theta}_{B}$, and $\bm{\Theta}_{D}$, as detailed respectively in subfigures (a), (b), and (d) of Fig.~\ref{fig:2d_subcri_sys_examples}. The 9 transient solutions converge along distinct trajectories toward three stationary distributions, $p_{\text{S}A}$, $p_{\text{S}B}$, and $p_{\text{S}D}$, attaching to the system parameters $\bm{\Theta}_{A}$, $\bm{\Theta}_{B}$, and $\bm{\Theta}_{D}$, respectively. The embeddings of the three transient solutions initialized at $p_{\text{I}A}$ (shown in cyan) follow a nearly straight path, reflecting the kind of behaviors illustrated in Fig.~\ref{fig:2d_subcri_sys_examples}(a), where probability diffuses isotropically from a peaked initial distribution toward the ring-shaped stationary distributions. In contrast, the embeddings of the six transient solutions started from $p_{\text{I}B}$ (shown in green) and $p_{\text{I}D}$ (shown in purple) move with rotation in the embedding space, which accurately captures the oscillatory dynamics seen in Figs.~\ref{fig:2d_subcri_sys_examples}(b) and (d) due to the steady increase of the phase angles on the probability masses.

Among the three stationary distribution embeddings, the embedding of $p_{\text{S}D}$ lies closest to the origin, while that of $p_{\text{S}A}$ is closer than that of $p_{\text{S}B}$. This pattern arises from the convexity-preserving nature of the embedding space introduced in Sec.~\ref{sec:gmd_ae_encoder} and the property that convex combinations of embeddings tend to shrink toward the origin, as formalized in Eq.~(\ref{eq:shrink}). The right-most column of Fig.~\ref{fig:2d_subcri_sys_examples} shows that, compared with $p_{\text{S}A}$, the distribution $p_{\text{S}D}$ is more uniform. Because such a nearly uniform distribution can be closely approximated by a mixture of many localized Gaussians, its convex combination in the embedding space is naturally located near the origin.

\begin{figure}[!htb]
\center{\includegraphics[width=1\textwidth]
{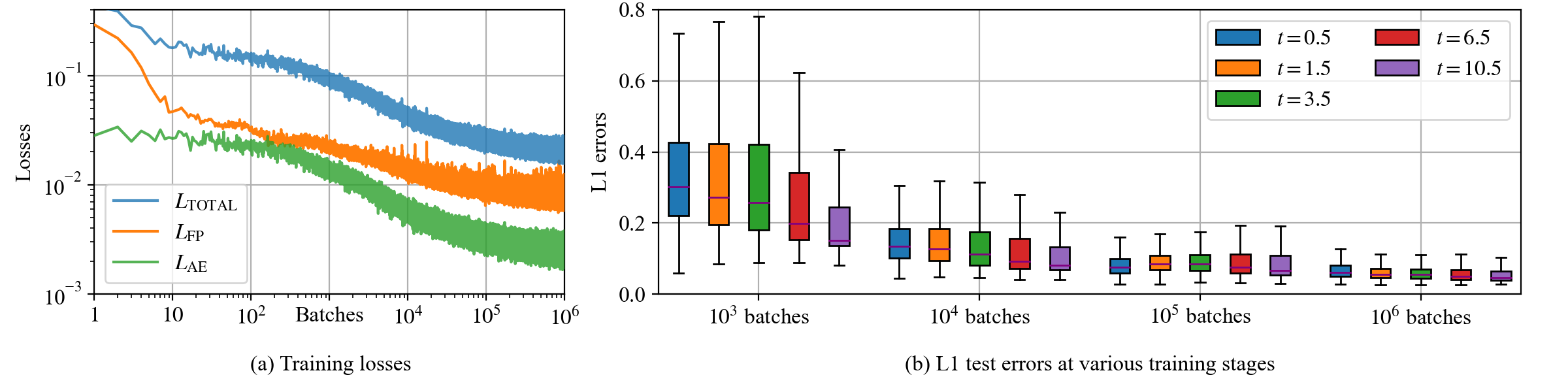}}
\caption{\label{fig:2d_subcritical_sys_loss} The training losses and test errors of the TPAPS on the 2-D subcritical system~(\ref{eq:subcri_sys}).}
\end{figure}

The training of the TPAPS on the 2-D subcritical system~(\ref{eq:subcri_sys}) is stable, as the training loss curves shown in Fig.~\ref{fig:2d_subcritical_sys_loss}(a) decrease steadily. The L1 errors between the transient distributions of the TPAPS and the MCS are detailed in Fig.~\ref{fig:2d_subcritical_sys_loss}(b) and Tab.~\ref{tab:error_2D_subcri}. The 2,500 choices of system parameters in the parameter domain $\mathcal{P}$ and initial distributions in IGMS $\widehat{\mathcal{K}}$ are randomly chosen. Each transient solution by the MCS is calculated on $3\times10^7$ trajectories and quantified into a $200\times 200$ grid in the square domain $[-2, 2]\times[-2, 2]$. The L1 error of the TPAPS decreases when the training continues at all the times, as shown in Tab.~\ref{tab:error_2D_subcri}. The time $t=10.5$ in the last row of Tab.~\ref{tab:error_2D_subcri} is much longer than the expected time $T_\text{max}=7$ for accurate prediction. This observation indicates that the TPAPS successfully generalizes to long-term near-stationary distributions. 

The training speed of the TPAPS is 0.90 seconds per batch on a system equipped with an Intel i9-13900K CPU and an NVIDIA GeForce RTX 4090 GPU and $10^5$ batches required 25 hours. In the test stage, it takes 8.3 seconds to evaluate 2,500 transient solutions with random initial distributions and control parameters. Here, each solutions include 10 snapshots with $t\in[0, 12]$ and each snapshot includes a $200\times 200$ grid of states. 3.7 seconds are used for calculating the transient GMDs and 4.6 seconds are spent on evaluating the 2-D grid of densities. In contrast, the MCS with $3\times10^7$ trajectories requires 293.4 seconds on the GPU or around $2.9\times10^4$ seconds on the CPU to calculate a single transient solution. Therefore, the TPAPS is around $8.8\times10^4$ times faster than the GPU-accelerated MCS at the test stage.

\begin{table}[!htb]
\footnotesize
\centering
\caption{The mean, SD, and median (MEAN/SD/MED) of the L1 errors between TPAPS and MCS results across 2,500 random transient distributions of the 2-D subcritical system~(\ref{eq:subcri_sys}) at various training stages.}\label{tab:error_2D_subcri}
\begin{tabular}{ccccc}
\toprule
Time & $10^3$ batches & $10^4$ batches & $10^5$ batches & $10^6$ batches\\
\midrule
$t=0.5$ & 0.3442/0.1681/0.3007 & 0.1501/0.0679/0.1348 & 0.0846/0.0365/0.0750 & 0.0689/0.0302/0.0610\\
$t=1.5$ & 0.3393/0.2051/0.2731 & 0.1535/0.0891/0.1273 & 0.0954/0.0451/0.0848 & 0.0661/0.0365/0.0558\\
$t=3.5$ & 0.3417/0.2314/0.2566 & 0.1511/0.1115/0.1119 & 0.0985/0.0535/0.0842 & 0.0670/0.0438/0.0547\\
$t=6.5$ & 0.2978/0.2333/0.1985 & 0.1444/0.1347/0.0912 & 0.1000/0.0712/0.0748 & 0.0662/0.0503/0.0503\\
$t=10.5$ & 0.2439/0.2101/0.1502 & 0.1431/0.1650/0.0801 & 0.1036/0.0926/0.0666 & 0.0649/0.0551/0.0462\\
\bottomrule
\end{tabular}
\end{table}

\begin{figure}[!htb]
\center{\includegraphics[width=1\textwidth]
{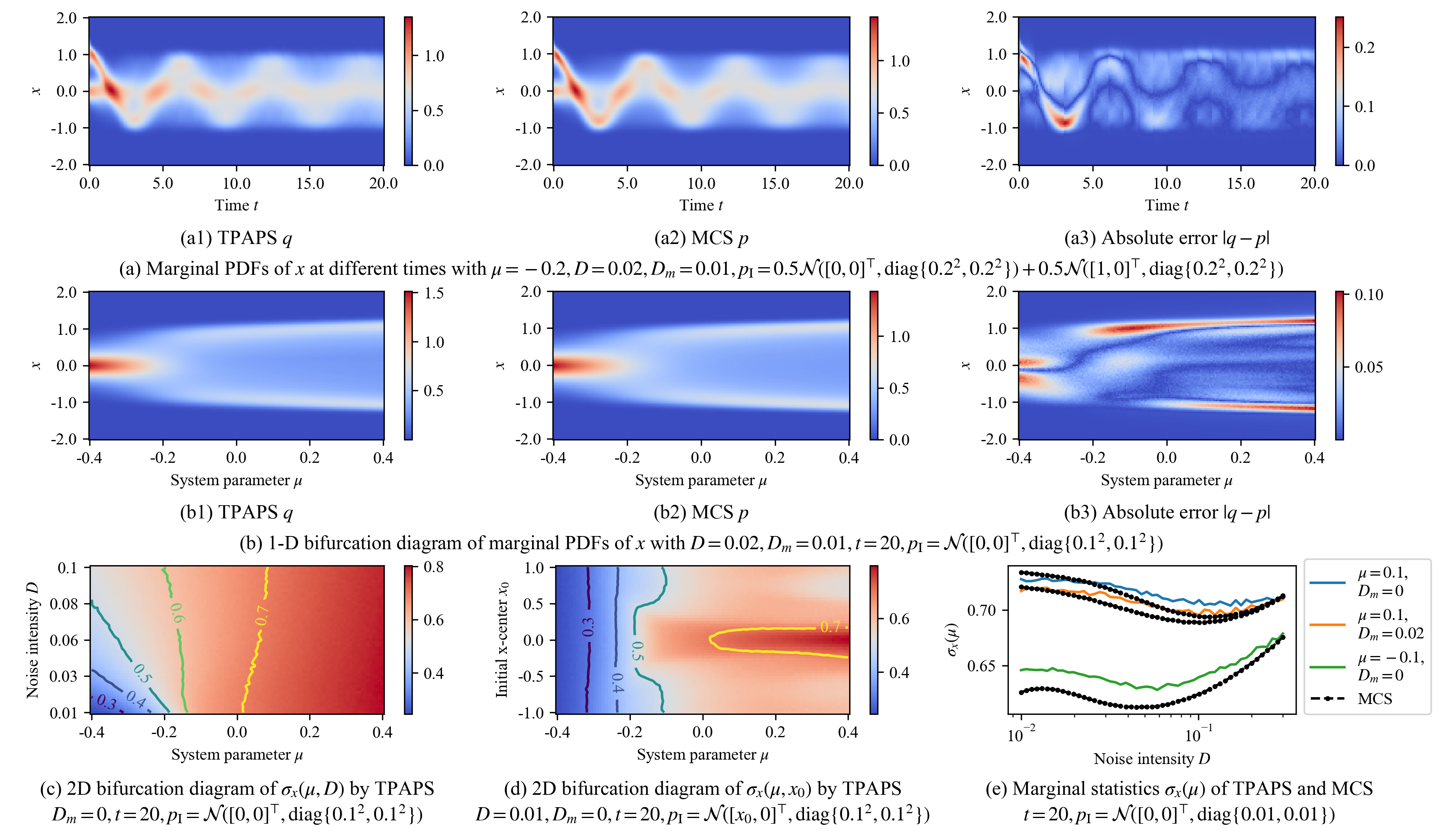}}
\caption{\label{fig:2d_subcritical_sys_examples}Efficient study of transient and near-stationary ($t=20$) dynamics of the 2-D subcritcal system~(\ref{eq:subcri_sys}) using the TPAPS. (a) shows an example of marginal transient distributions of $x$ at different times. (b) shows the Hopf bifurcation of stationary distributions triggered by the system parameter $\mu$ by visualizing the marginal stationary distributions of $x$. (c) investigates two-parameter bifurcation of stationary distributions depending on noise intensity $D$ and system parameter $\mu$ in terms of the SD $\sigma_x$ of the marginal distribution of $x$ and (d) demonstrates the two-parameter bifurcation of the non-ergodic distributions depending on $\mu$ and the x-coordinate $x_0$ of the center of initial distributions. (e) compares several groups of the SD $\sigma_x$ of the distributions obtained by TPAPS and the MCS.}
\end{figure}

The core strength of TPAPS lies in its ability to rapidly generate large numbers of transient distributions in all the space $\mathcal{S}\times\mathcal{T}\times\mathcal{P}\times\mathcal{K}$. The stationary distributions can be approximated by selecting a large time, e.g., $t=20$. This allows us to efficiently explore the transient and stationary responses of stochastic systems at different time points under varying system parameters and initial distributions, without the need for time-consuming repeated simulations. Four examples are gathered in Fig.~\ref{fig:2d_subcritical_sys_examples}, where the subfigures have a resolution of $200 \times 100$ grid points along the horizontal and vertical axes, respectively. The TPAPS results in Fig.~\ref{fig:2d_subcritical_sys_examples}(a1)(b1)(c) and (d) take 0.07, 0.07, 8.1, and 7.6 seconds, respectively, showing that it supports real-time multi-parameter bifurcation study.

Figure~\ref{fig:2d_subcritical_sys_examples}(a1) illustrates a series of marginal transient distributions of $x\in[-2, 2]$ over the 200 times $t$ equally spaced in $[0,20]$, obtained by omitting the $y$-component from the TPAPS $q$ and evaluating the marginal GMD of $x$ using Eq.~(\ref{eq:gaussian_mixture}). The initial distribution is $0.5\mathcal{N}([0, 0]^\top,\text{diag}\{0.04, 0.04\})+0.5\mathcal{N}([1, 0]^\top,\text{diag}\{0.04, 0.04\})$. The TPAPS restores both the oscillations due the initial Gaussian component centered at $[1,0]^\top$ and the persistent probability density near the origin due to the initial Gaussian component centered at the origin. These marginal distributions are compared with those from the MCS $q$ with $3 \times 10^6$ trajectories in Fig.~\ref{fig:2d_subcritical_sys_examples}(a2). The TPAPS approximates the MCS for both short-term and long-term distributions. Figure~\ref{fig:2d_subcritical_sys_examples}(a3) reveals a distinct mosaic pattern in the absolute errors $|q-p|$ between TPAPS and MCS whenever $t$ is an integer multiple of the leap time $t_\text{LEAP}=1$. This pattern arises directly from the recursive scheme in Eq.~(\ref{eq:TPAPS_full}) for ease of training and efficiency of computation, where predictions within each interval $(kt_\text{LEAP}, (k+1)t_\text{LEAP}]$ are computed after a differing number of recursive leaps, followed by a final short-term prediction.

Figure~\ref{fig:2d_subcritical_sys_examples}(b) captures the subcritical Hopf bifurcation in the 2-D system~(\ref{eq:subcri_sys}) by displaying the marginal near-stationary distributions of $x\in[-2, 2]$  from both TPAPS $q$ and MCS $p$ at a large time $t=20$. A unimodal distribution centered at the origin persists for $\mu < -0.25$, which vanishes once $\mu>-0.25$, giving way to a LCO. The consistently small absolute errors in Fig.~\ref{fig:2d_subcritical_sys_examples}(b3) compared with the MCS in Fig.~\ref{fig:2d_subcritical_sys_examples}(b2) demonstrate the effectiveness of the TPAPS approximation throughout this characteristic transition.

Two 2-D bifurcation diagrams are numerically investigated via the TPAPS in Figs.~\ref{fig:2d_subcritical_sys_examples}(c) and (d). The stochasticity of the near-stationary distribution of the system at $t=20$ is quantified by the sample SD $\sigma_x$ of the marginal distribution of $x$, computed from 50,000 random samples drawn from the GMD provided by the TPAPS. Note that the GMD generated by TPAPS may include a few components with very low weights yet very large variances. While these components contribute negligibly to the overall L1 error, they can significantly skew the estimated variance of the transient distribution. To eliminate this statistical artifact and ensure an accurate marginal SD, rare samples lying outside the state domain $\mathcal{S}$ are filtered out. Figures~\ref{fig:2d_subcritical_sys_examples}(c) demonstrates the variation of $\sigma_x$ conditioned on the system parameter $\mu$ and the noise intensity $D$. When $D$ increases, the stationary distributions exhibit greater randomness, as quantified by their increasing variances. Increasing $\mu$ also leads to larger variances, as the system undergoes a bifurcation from a stable fixed point to a large-amplitude LCO. 

Figures~\ref{fig:2d_subcritical_sys_examples}(d) detects the ergodicity of the system~(\ref{eq:subcri_sys}) by examining how the near-stationary distributions depend on the initial condition $x_0$ for different values of the bifurcation parameter $\mu$. The initial distribution is a Gaussian $\mathcal{N}([x_0,0]^\top,\text{diag}\{0.01, 0.01\})$. When $\mu<-0.2$, the system possesses a unique stationary distribution regardless of $x_0$, as the marginal SDs $\sigma_x$ are the same, indicating ergodic behavior. In contrast, for $\mu>-0.2$, the fixed point loses stability, the initial conditions near the origin quickly evolve to the LCO, whereas those starting far from the origin may undergo prolonged rotating transients before converging to the stationary isotropic cycle, leading to lower variances. This dependence on $x_0$ results in different effective stationary distributions within finite observation times, revealing the non-ergodic behavior of the system~(\ref{eq:subcri_sys}).

To quantify the accuracy of these numerical results, Fig.~\ref{fig:2d_subcritical_sys_examples}(e) compares the marginal SDs $\sigma_x$ from the TPAPS with the counterparts produced by the MCS. Three groups with different $\mu$ and noise intensity $D_m$ are considered. When $\mu=0.1, D_m=0$ or $\mu=0.1, D_m=0.02$, the marginal SDs by the TPAPS align closely to the MCS solutions at different values of the noise intensity $D$. When $\mu=-0.1,D_m=0$, the TPAPS solutions overestimate the marginal SDs slightly, though the overall shapes of the curves align well.

\subsection{Duffing system}
\label{sec:system_duffing}
Finally, we consider the second-order Duffing system~\cite{Xu2020Solving},
\begin{equation}
\ddot{x} + \eta \dot{x} + \alpha x + \beta x^3 = \sigma \xi(t),\nonumber
\end{equation}
with the damping parameter $\eta$, the stiffness parameters $\alpha$ and $\beta$, and the noise amplitude $\sigma$. By defining the 2-D state vector $\mathbf{x}(t)=[x(t), \dot{x}(t)]^\top\triangleq [x(t), y(t)]^\top$, the system can be reformulated as first-order SDEs
\begin{equation}\label{eq:duffing}
\begin{split}
\dot{x}&=y\\
\dot{y}&=-\eta y - (\alpha x + \beta x^3) + \sigma \xi(t).
\end{split}
\end{equation}
The corresponding FP operator of the transient distribution $p=p(\mathbf{x},t;\bm{\Theta}, p_\text{I})$ with 4 parameters $\bm{\Theta}=[\eta, \alpha, \beta, \sigma]^\top$ is
\begin{equation}
\mathcal{L}_\text{FP}p=-\frac{\partial (yp)}{\partial x} - \frac{\partial}{\partial y}\left[-(\eta y + \alpha x + \beta x^3)p\right]+\frac{\sigma^2}{2}\frac{\partial^2 p}{\partial x^2}.\nonumber
\end{equation}
The Duffing system also has the analytical stationary distribution
\begin{equation}\label{eq:p_s_duffing}
p_\text{S}(\mathbf{x};\bm{\Theta})=C\cdot \text{exp}\left[-\frac{\eta}{\sigma^2}\left(\alpha x^2+\frac{\beta x^4}{2}+y^2\right)\right],
\end{equation}
where the normalization factor $C$ is calculated by numerical integration.

We select the 2-D state domain $\mathcal{S}=\{\mathbf{x}=[x, y]^\top|x,y\in[-5,5]\}$ and the 4-D parameter domain $\mathcal{P}=\{\eta, \alpha,\beta, \sigma\in[0.2, 1]\}$. The IGMS $\widehat{\mathcal{K}}$ in Eq.~(\ref{eq:IGMS}) includes GMDs with $N_\text{INIT}=5$ 2-D Gaussian components with means $\mu_{jk}\in[-2, 2]$ and standard deviations $\sigma_{jk}\in[0.1,0.5]$ for $j=1,\cdots,5$ and $k=1,2$. A TPAPS is trained for $10^6$ batches and the sizes of samples in each batch are $B_\text{FP}=150$, $B_\text{AE}=1500$, $N_\text{FP}=200$, $N_\text{AE}=2000$, $N_\text{NORM}=200$. The time leap is $t_\text{LEAP}=3$ and the maximal initial time is $T_\text{INIT}=9$. The expected time horizon for accurate prediction is $T_\text{max}=T_\text{INIT}+t_\text{LEAP}=12$.

\begin{figure}[!htb]
\center{\includegraphics[width=1\textwidth]
{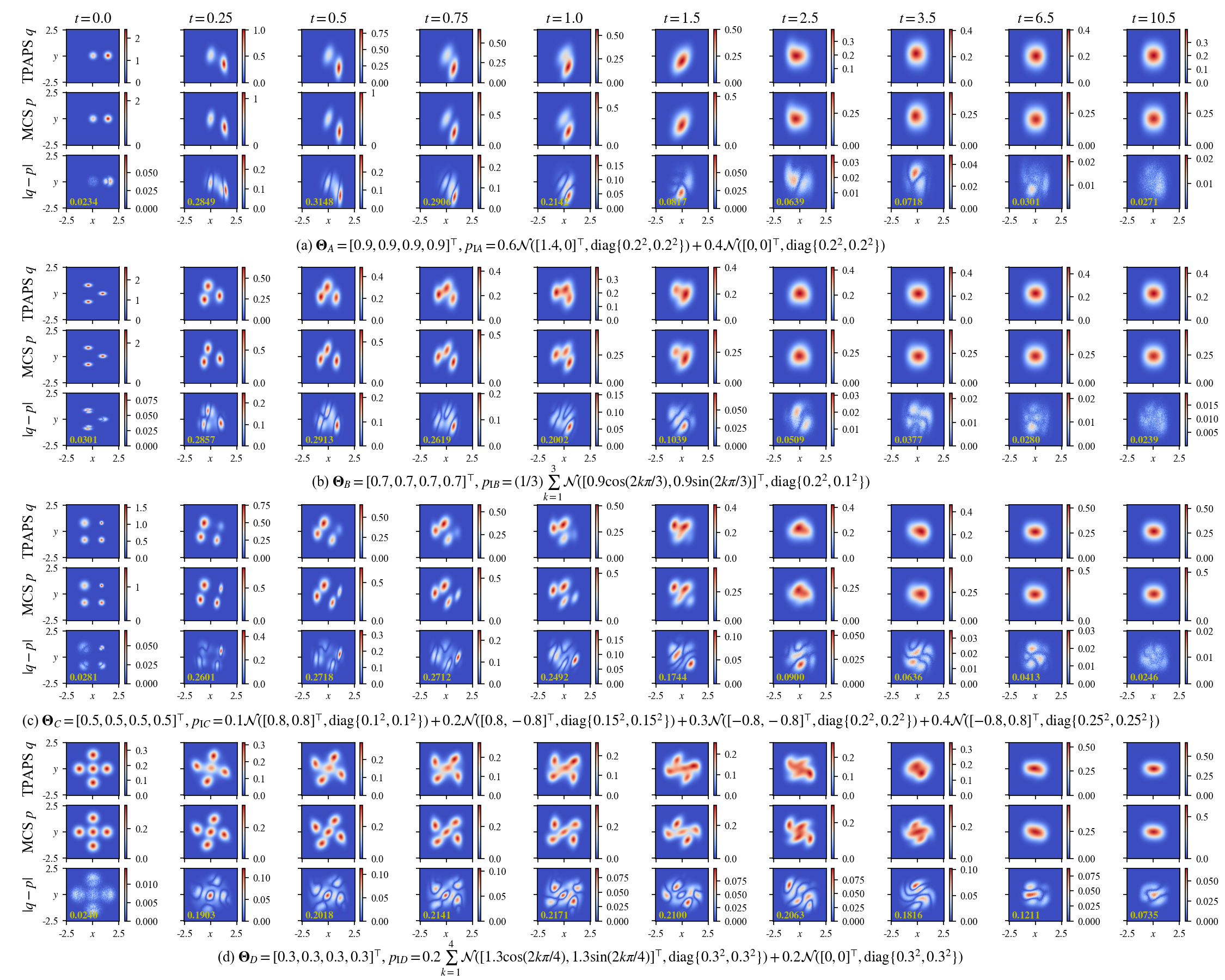}}
\caption{\label{fig:2d_duffing_sys_examples}Four transient solutions of the 2-D Duffing system~(\ref{eq:duffing}) by TPAPS and MCS with different initial distributions and system parameters. The L1 errors are displayed on the residual images.}
\end{figure}

In Fig.~\ref{fig:2d_duffing_sys_examples}, the TPAPS is applied to solve four transient solutions with different system parameters and complex initial distributions. The comparisons with MCS are also drawn. The initial distributions $p_{\text{S}A}$, $p_{\text{S}B}$, $p_{\text{S}C}$, and $p_{\text{S}D}$ have 2 to 5 Gaussian components with possibly different variances. The first column ($t=0$) of Fig.~\ref{fig:2d_duffing_sys_examples} indicates that the GMD autoencoder learns the initial distributions well. The last column ($t=10.5$) shows that in all cases the systems attain the corresponding stationary distributions, which are predicted by the TPAPS with high precision. The middle columns show that the transient distributions with several clusters can be faithfully predicted by the TPAPS, demonstrating its ability to jointly predict transient dynamics across diverse control parameters and initial conditions. Nevertheless, the TPAPS and MCS are not perfectly matched. The TPAPS fails in precisely capturing the rightmost probability mass at $t=0.25, 0.5, 0.75$ in Fig.~\ref{fig:2d_duffing_sys_examples}(c) and the five close probability components in Fig.~\ref{fig:2d_duffing_sys_examples}(d).

\begin{figure}[!htb]
\center{\includegraphics[width=1\textwidth]
{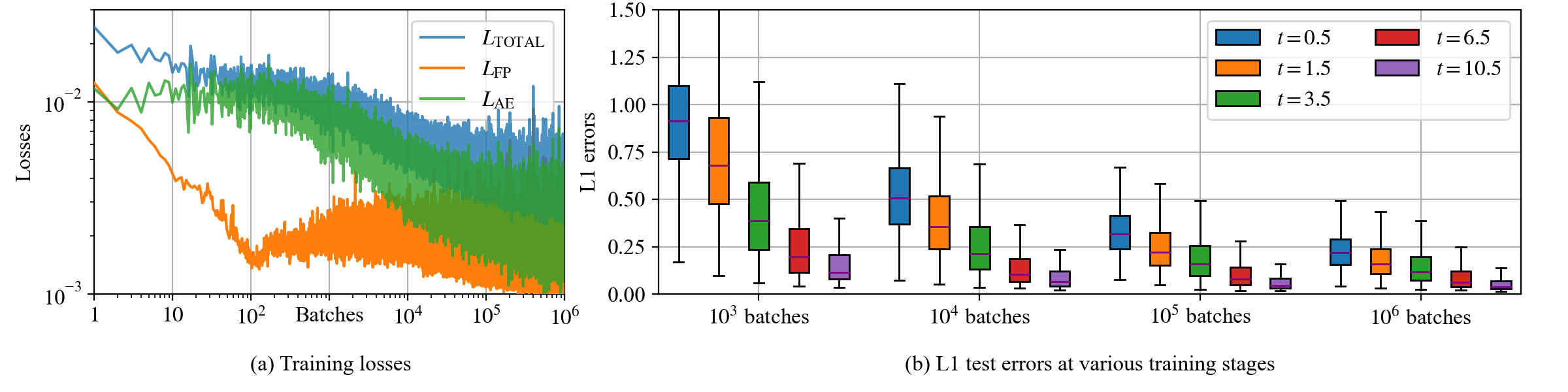}}
\caption{\label{fig:2d_duffing_sys_loss}The training losses and test errors of the TPAPS on the 2-D Duffing system~(\ref{eq:duffing}).}
\end{figure}

Figure~\ref{fig:2d_duffing_sys_loss}(a) details the loss curves of TPAPS for training the Duffing system~(\ref{eq:duffing}). The FPE loss $L_\text{FP}$ drops rapidly below 0.002 within the first 100 training batches and remains around this level thereafter. In contrast, the autoencoder loss $L_\text{AE}$ decreases steadily throughout the entire training process. This discrepancy indicates that the EVO-RN learns very quickly, enabling it to generate transient distributions that consistently yield low FPE residuals. However, these distributions may not correspond to the true transient states of the Duffing system~(\ref{eq:duffing}), as the autoencoder learns at a much slower pace and reconstructs them with a low accuracy.

\begin{table}[!htb]
\footnotesize
\centering
\caption{The mean, SD, and median (MEAN/SD/MED) of the L1 errors between TPAPS and MCS results across 2,500 random transient distributions of the Duffing system~(\ref{eq:duffing}) at various training stages.}\label{tab:error_duffing}
\begin{tabular}{ccccc}
\toprule
Time & $10^3$ batches & $10^4$ batches & $10^5$ batches & $10^6$ batches\\
\midrule
$t=0.5$ & 0.9115/0.2776/0.9150 & 0.5292/0.2185/0.5053 & 0.3340/0.1317/0.3168 & 0.2331/0.1051/0.2166\\
$t=1.5$ & 0.7175/0.3149/0.6793 & 0.4036/0.2224/0.3551 & 0.2549/0.1435/0.2194 & 0.1892/0.1140/0.1589\\
$t=3.5$ & 0.4466/0.2799/0.3853 & 0.2747/0.2043/0.2143 & 0.1967/0.1413/0.1585 & 0.1524/0.1159/0.1160\\
$t=6.5$ & 0.2624/0.2023/0.1971 & 0.1528/0.1415/0.1027 & 0.1139/0.0995/0.0810 & 0.0963/0.0950/0.0630\\
$t=10.5$ & 0.1693/0.1353/0.1152 & 0.0969/0.0860/0.0648 & 0.0666/0.0586/0.0457 & 0.0600/0.0623/0.0364\\
\bottomrule
\end{tabular}
\end{table}

The prediction of the TPAPS on the Duffing system~(\ref{eq:duffing}) is quantitatively evaluated by randomly sampling 2,500 initial distributions in the IGMS $\widehat{\mathcal{K}}$ and system parameters in the parameter domain $\mathcal{P}$ and comparing the L1 distances between the transient solutions of TPAPS and MCS. Several boxplots are shown in Fig.~\ref{fig:2d_duffing_sys_loss}(b) and key statistics are summarized in Tab.~\ref{tab:error_duffing}. The main pattern is that the prediction errors at all times $t$ decrease when the training batches increase, indicating that the learning is stable. 

\begin{figure}[!htb]
\center{\includegraphics[width=1\textwidth]
{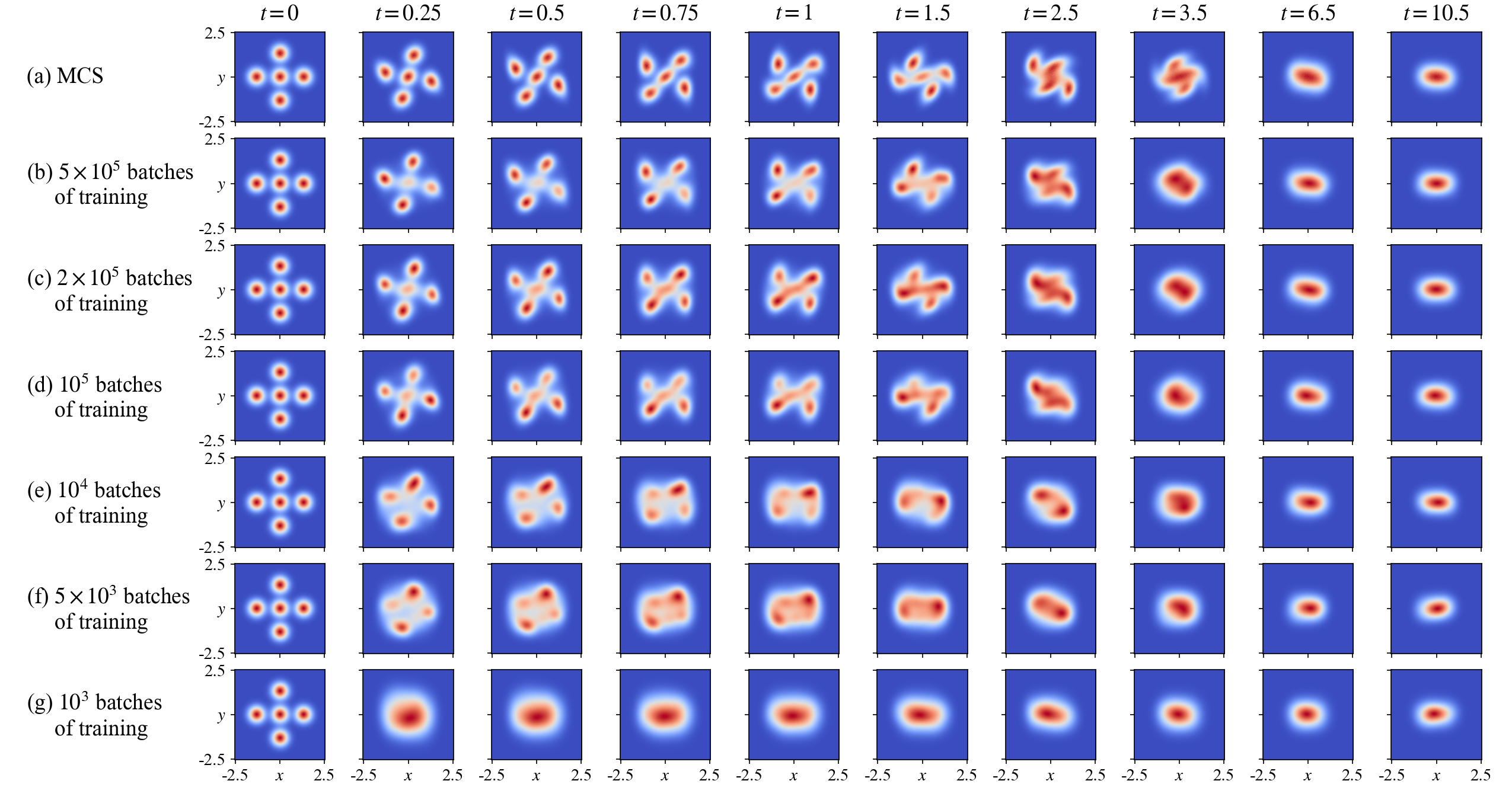}}
\caption{\label{fig:2d_duffing_diff_batches}The TPAPS results of the Duffing system~(\ref{eq:duffing}) at early training stages. The initial distribution, system parameters, and the results at $10^6$ training batches have been detailed in Fig.~\ref{fig:2d_duffing_sys_examples}(d).}
\end{figure}

Figure~\ref{fig:2d_duffing_diff_batches} details the prediction of TPAPS at various training stages, preceding to the final $10^6$ batches shown in Fig.~\ref{fig:2d_duffing_sys_examples}(d). When the training continues, the short-term distributions, such as $t=0.25$, are more accurate, enforcing the improvement of the distributions at larger times. Another distinct pattern in both Figs.~\ref{fig:2d_duffing_sys_loss}(b) and \ref{fig:2d_duffing_diff_batches} is that the errors at the time $t=10.5$, which correspond to the stationary regime, are already small after only 1,000 training batches. It suggests that early in training, when the autoencoder cannot yet generate reliable short-term transient distributions, the EVO-RN prioritizes learning the stationary dynamics first. In contrast, accurately capturing the short-term dynamics for $t<6.5$ is more challenging. Because the EVO-RN must rely on the slow learning autoencoder to first reconstruct faithful transient distributions before it can effectively model the finer temporal evolution. This again justifies assigning more samples and a higher penalty to the autoencoder loss than to the FPE loss.

The training speed of the TPAPS is 1.11 seconds per batch and $10^5$ batches needs 30.8 hours. At the test stage, it takes 0.95 second to recursively obtain the GMDs of 2,500 transient solutions and the calculation of densities at states on a $200\times 200$ grid takes 24.1 seconds. Here, each transient solution includes 10 snapshots with different $t\in [0, 12]$. The MCS with $3\times10^7$ trajectories for calculating one such transient distribution requires 93.2 seconds on the GPU or around $9.6\times10^3$ seconds on the CPU.

\begin{figure}[!htb]
\center{\includegraphics[width=1\textwidth]
{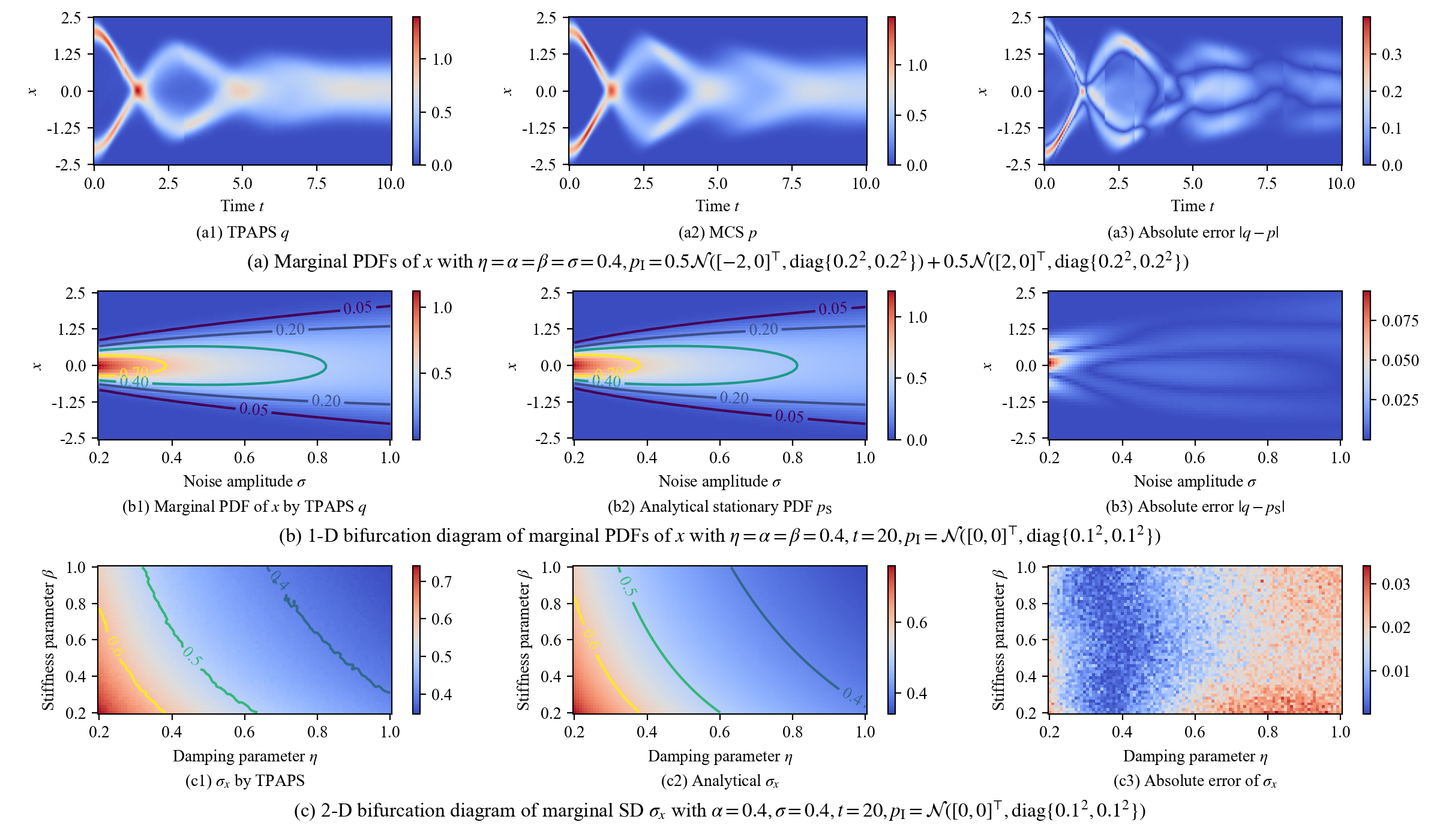}}
\caption{\label{fig:2d_duffing_sys_bifu}Efficient study of transient and near-stationary ($t=20$) dynamics of the Duffing system~(\ref{eq:duffing}) using the TPAPS. (a) shows marginal transient distributions of $x$ at different times. (b) shows the impact of the noise amplitude $\sigma$ on the marginal stationary distribution of $x$ at $t=20$. (c) investigates the dynamics of stationary distributions depending on the damping parameters $\eta$ and the stiffness parameter $\beta$. The true transient solutions are obtained by MCS in (a2) and the true stationary results in (b2) and (c2) are calculated by Eq.~(\ref{eq:p_s_duffing}). The resolution of all subfigures is $200 \times 100$.}
\end{figure}

Figure~\ref{fig:2d_duffing_sys_bifu} investigates several aspects of the Duffing system~(\ref{eq:duffing}) using the TPAPS and compares it with the transient distributions obtained by the MCS or the analytical stationary distributions by Eq.~(\ref{eq:p_s_duffing}). The TPAPS results in Figs.~\ref{fig:2d_duffing_sys_bifu}(a1), (b1), and (c1) only take 0.02, 0.04, and 7.4 seconds on GPU, respectively.

Figure~\ref{fig:2d_duffing_sys_bifu}(a) shows an example of the marginal transient distributions of $x$ at $t\in[0, 10]$. The two equally weighted Gaussian components centered at $[2,0]^\top$ and $[-2,0]^\top$ in the initial distribution spirally evolve to the stationary distribution centered at the origin, leading to the projected crosses at $x=0$. The MCS in Fig.~\ref{fig:2d_duffing_sys_bifu}(a2) is calculated on $3\times 10^7$ trajectories. The TPAPS accurately captures this complex transient dynamics as Fig.~\ref{fig:2d_duffing_sys_bifu}(a3) shows that the errors compared with the MCS are relatively low.

The near-stationary dynamics of the Duffing system~(\ref{eq:duffing}) is explored by choosing a large time $t=20$ in the TPAPS. Figure~\ref{fig:2d_duffing_sys_bifu}(b1) demonstrates the marginal stationary distribution of $x$ by the TPAPS with different noise amplitude $\sigma$. It can be seen that a larger $\sigma$ results in a larger variance of $x$. Compared with the true analytical solutions in Fig.~\ref{fig:2d_duffing_sys_bifu}(b2), the TPAPS obtains accurate stationary approximations. Figure~\ref{fig:2d_duffing_sys_bifu}(c) further explores the stationary behaviors of the Duffing system conditioned on the damping parameter $\eta$ and the stiffness parameter $\beta$. The transient GMDs at the large time $t=20$ are obtained by the TPAPS and the sample SD $\sigma_x$ of the marginal distributions of $x$ are calculated and shown in Fig.~\ref{fig:2d_duffing_sys_bifu}(c1). The TPAPS reproduces the physical finding that increased damping or stiffness suppresses stochasticity, resulting in lower variance, consistent with theoretical expectation. The statistics calculated by the TPAPS are compared with the analytical solutions in Fig.~\ref{fig:2d_duffing_sys_bifu}(c2). The low absolute errors in Fig.~\ref{fig:2d_duffing_sys_bifu}(c3) indicate the high accuracy of our method in producing characterizing statistics of long-term responses.

\section{Discussion}
\label{sec:discussion}
To investigate the hyperparameters of the TPAPS, we conduct an ablation study on the Duffing system~(\ref{eq:duffing}), using the original configuration as the baseline. Three variants (TPAPS-A, B, C) are evaluated, with each model trained for $10^6$ batches. Table~\ref{tab:ablation} lists the average L1 errors computed on 2500 transient distributions in producing Tab.~\ref{tab:error_duffing}, recorded at $10^5$ and $10^6$ training batches.

\begin{table}[!htb]
\tiny  
\centering
\caption{The mean L1 errors between several TPAPS and MCS results across 2,500 random transient distributions of the Duffing system~(\ref{eq:duffing}) at different times, while the last column shows the training speed. Best values are bold.}\label{tab:ablation}
\begin{tabular}{lcccccccccccc}
\toprule
\multirow{2}{*}{Method} & \multicolumn{5}{c}{$10^5$ training batches} & & \multicolumn{5}{c}{$10^6$ training batches} & \multirow{2}{*}{seconds/batch}\\
\cline{2-6}\cline{8-12} & $t=0.5$ & $t=1.5$ & $t=3.5$ & $t=6.5$ & $t=10.5$ & & $t=0.5$ & $t=1.5$ & $t=3.5$ & $t=6.5$ & $t=10.5$ &\\
\midrule
Baseline & 0.3340 & 0.2549 & 0.1967 & 0.1139 & 0.0666 && 0.2331 & $\mathbf{0.1892}$ & 0.1524 & 0.0963 & 0.0600 & 1.11\\
TPAPS-A & $\mathbf{0.2296}$ & $\mathbf{0.2290}$ & $\mathbf{0.1768}$ & 0.1301 & 0.1098 && $\mathbf{0.1896}$ & 0.1994 & $\mathbf{0.1459}$ & 0.1204 & 0.1037 & 1.16\\
TPAPS-B & 0.3263 & 0.2598 & 0.1968 & 0.1098 & 0.0613 && 0.2532 & 0.2035 & 0.1616 & 0.0952 & 0.0541 & 0.48\\
TPAPS-C & 0.3402 & 0.2695 & 0.2114 & 0.1340 & 0.0781 && 0.2875 & 0.2275 & 0.1695 & 0.0996 & 0.0596 & 0.67\\
TPAPS-D & 0.3583 & 0.2835 & 0.2007 & $\mathbf{0.1067}$ & $\mathbf{0.0547}$ && 0.2910 & 0.2261 & 0.1606 & $\mathbf{0.0845}$ & $\mathbf{0.0446}$ & 1.08\\

\bottomrule
\end{tabular}
\end{table}

In TPAPS-A, the leap time $t_\text{LEAP}$ is reduced from 3 to 1, and the time horizon of the TGMS $t_\text{INIT}$ from 9 to 6. The smaller leap time reduces short-term errors at $t=0.5$ and 1.5. However, long-term errors for $t>6$ become significantly higher than the baseline. These results indicate that a small leap time can improve short-term accuracy, but also suggest that the parameter $t_\text{INIT}$ should be sufficiently long, i.e., comparable to the time for the system to reach stability, so that the TGMS can capture the stationary dynamics.

TPAPS-B features a shallower GMD autoencoder. The numbers of residual blocks in the EMB-RN, REP-RN, and DEC-RN are reduced from 3, 3, and 6 to 1, 1, and 2, respectively. This architecture reduction saves 57\% of the training time, as shown in the last column of Tab.~\ref{tab:ablation}. However, accuracy is compromised and becomes poor in most cases. This confirms that sufficient depth in the GMD autoencoder is crucial for the accuracy of TPAPS.

In TPAPS-C, the number of Gaussian components $N_\text{GAU}$ in the reconstructed GMD is reduced from 100 to 50. This simplification saves 25\% of the training time but yields a considerably low accuracy, especially when $t\leq 3.5$. This demonstrates that a sufficient number of Gaussian components is essential for the representation capability of the GMD, particularly when modeling evolving distributions that contain multiple high-probability clusters.

In TPAPS-D, the sample fractions $\lambda_\text{AE}$ and $\lambda_\text{FP}$ are reduced from 0.75 to 0.25. As a result, for calculating the autoencoder and FPE losses, only 25\% of the distributions are sampled from the IGMS and 75\% from the TGMS. This change yields the highest accuracy for predicting long-term dynamics ($t \ge 6.5$). However, it also results in the worst performance for short-term dynamics ($t \le 1.5$), indicating a trade-off between short-term and long-term predictive accuracy. Since a key goal for TPAPS is to accurately predict short-term transient dynamics, this result validates the rationale behind choosing $\lambda_\text{AE} = \lambda_\text{FP} = 0.75$.

The main bottleneck of the TPAPS is the slow training speed due to the vast variations of the nonlinear dynamics rooted in the variable initial condition, evolution time, and system parameters. A potential direction for future improvement is to increase sampling efficacy, for example, through adaptive resampling~\cite{Zhang2026Solving} or a mixture of uniform and importance sampling~\cite{Li2023Artificial}. Potential speedup strategies also include developing a universal, capacious GMD encoder applicable to all systems of a given dimensionality, or reusing pre-trained encoders from systems with similar dynamics via transfer learning~\cite{Wang2025transfer}.

\section{Conclusion}
\label{sec:conclusion}
This work introduced a novel end-to-end PAPS for the parallel solving of transient FPEs. By integrating a constraint-preserving autoencoder for PDFs with a neural network that learns latent-space dynamics and a recursive time-leaping scheme, the proposed method achieved a single unified model capable of generating transient PDFs across diverse initial conditions, system parameters, and time instants in a single forward pass. The core innovation lies in a modular paradigm that decouples representation learning from physics, where an autoencoder constructs a universal, low-dimensional manifold for representing GMDs, while a separate EVO-RN learns the pure FPE dynamics within this structured latent space. This decomposition transforms the intractable challenge of parallel FPE solving into two manageable sub-problems. Crucially, the framework inherently preserves the mathematical constraints of PDFs and through its specialized architecture, such as normalization, non-negativity, and variable initial condition, eliminating the need for complex penalty terms in the loss function. 

Extensive numerical experiments demonstrated that the proposed method can solve the FPE in parallel for both short-term transient and long-term stationary solutions, even under challenging multi-modal initial distributions and across varying system parameters. The algorithm's inference speed remains four orders of magnitude faster than GPU-accelerated MCS and six orders faster than the traditional CPU-based MCS. This efficiency enables real-time, high-throughput numerical studies of stochastic bifurcation, both in single- and multi-parameter spaces, such as analyzing the influence of parameters on stationary distributions in ergodic systems or investigating initial-condition dependence in non-ergodic systems. Consequently, the method provides a powerful tool for exhaustively exploring the transient dynamics of multi-dimensional, multi-parameter, nonlinear stochastic systems across their parameter spaces. A current limitation of the algorithm is its relatively slow training speed. Future work will focus on improving the sampling strategy and investigating the method's applicability to higher-dimensional systems.

\appendix
\section{Euler-Maruyama method}
\label{sec:euler_maruyama}
To obtain the true transient solution of the FPE in Eq.~(\ref{eq:fpe}) for comparison, the system~(\ref{eq:sde}) is discretized by the Euler-Maruyama scheme
\begin{equation}
\mathbf{x}_{n+1} = \mathbf{x}_n + \delta t \cdot \mathbf{A}(\mathbf{x}_n, t_n;\bm{\Theta}) + \sqrt{\delta t} \cdot \mathbf{B}(\mathbf{x}_n, t_n;\bm{\Theta}) \cdot [\zeta_{n1}, \cdots, \zeta_{nM}]^\top,
\end{equation}
where $n$ is the iteration number, $\delta t$ is the discretization time step such that $t_n=n \cdot \delta t$, $\mathbf{x}_n=\mathbf{x}(t_n)$, and $\{\zeta_{ni}\}_{i=1}^M$ are independent standard Gaussian random numbers. The time step $\delta t$ is set to 0.001 in all experiments. The initial state $\mathbf{x}_0$ is drawn from the initial distribution $p_\text{I}(\mathbf{x})$ and repeated for $N$ times, leading to $N$ trajectories. 

This method is implemented in PyTorch and fully vectorized, enabling simultaneous evolution of $N$ trajectories. We adopt in-place operations, storing only the essential state information and the histograms or statistics required for simulation, thereby reducing the memory footprint to $O(N)$. The simulation can be executed on CPU and significantly accelerated via parallel computation on GPU.

\textbf{Remark}: Unlike GPU-accelerated MCS, which still requires a GPU and repeated simulations per query, our TPAPS enables real-time inference on a CPU after a single training session, making large-scale parameter sweeps feasible even without dedicated hardware.

\section*{CRediT authorship contribution statement}
Xiaolong Wang: Conceptualization, Methodology, Software, Writing - original draft. Jing Feng: Formal analysis, Writing - original draft, Writing - review \& editing. Qi Liu: Writing - original draft, Writing - review \& editing. Chengli Tan: Writing - original draft, Writing - review \& editing. Yuanyuan Liu: Writing - original draft, Writing - review \& editing. Yong Xu: Conceptualization, Writing - original draft, Writing - review \& editing.

\section*{Funding}
This study was partly supported by the NSF of China (Grant No. 52225211), the Key International (Regional) Joint Research Program of the NSF of China (Grant No. 12120101002), and the National Natural Science Foundation of China (Grant Nos. 12202255, 12102341, and 12072264).

\section*{Declaration of competing interest}
The authors declare that they have no known competing financial interests or personal relationships that could have appeared to influence the work reported in this paper.

\section*{Data availability}
Data will be made available on request.

\normalem
\bibliographystyle{plain}  
\bibliography{references} 

\end{document}